\documentclass[]{aa} 

\usepackage{graphicx}
\usepackage{txfonts}
\usepackage{natbib}
\usepackage{longtable}
\usepackage{lscape}

\begin{document}

\title{Galactic cold cores
III. General cloud properties
\thanks{
{\it Planck} \emph{(http://www.esa.int/Planck)} is a project of the European Space
Agency -- ESA -- with instruments provided by two scientific consortia funded
by ESA member states (in particular the lead countries: France and Italy) with
contributions from NASA (USA), and telescope reflectors provided in a
collaboration between ESA and a scientific Consortium led and funded by
Denmark.
}
\thanks{{\it Herschel} is an ESA space observatory with science instruments provided
by European-led Principal Investigator consortia and with important
participation from NASA.}
}

\author{M.     Juvela\inst{1},
        I.     Ristorcelli\inst{2,3},
        L.     Pagani\inst{4},        
        Y.     Doi\inst{6},
        V.-M.  Pelkonen\inst{1,7},
        D.J.   Marshall\inst{2,3},
        J.-P.  Bernard\inst{2,3},        
        E.     Falgarone\inst{5},
        J.     Malinen\inst{1},
        G.     Marton\inst{9},
        P.     McGehee\inst{7},
        L.A.   Montier\inst{2,3},
        F.     Motte\inst{8},
        R.     Paladini\inst{7},
        L.V.   T\'oth\inst{9},                
        N.     Ysard\inst{1,10},
        S.     Zahorecz\inst{9},
        A.     Zavagno\inst{11}
        }

\institute{
Department of Physics, P.O.Box 64, FI-00014, University of Helsinki,
Finland, {\em mika.juvela@helsinki.fi}                                
\and
Universit\'e de Toulouse, UPS-OMP, IRAP, F-31028 Toulouse cedex 4, France   
\and
CNRS, IRAP, 9 Av. colonel Roche, BP 44346, F-31028 Toulouse cedex 4, France  
\and
LERMA, CNRS UMR8112, Observatoire de Paris, 61 avenue de l'observatoire 75014 Paris, France
\and
LERMA, CNRS UMR8112, Observatoire de Paris and Ecole Normale Superieure,
24 rue Lhomond, 75005 Paris, France                                    
\and
The University of Tokyo, Komaba 3-8-1, Meguro, Tokyo, 153-8902, Japan 
\and
IPAC, Caltech, Pasadena, USA                                          
\and
Laboratoire AIM Paris-Saclay, CEA/DSM - INSU/CNRS - Universit\'e Paris Diderot,IRFU/SAp
CEA-Saclay, 91191 Gif-sur-Yvette, France                              
\and
Lor\'and E\"otv\"os University, Department of Astronomy,
P\'azm\'any P.s. 1/a, H-1117 Budapest, Hungary (OTKA K62304)          
\and
IAS, Universit\'e Paris-Sud, 91405 Orsay cedex, France                
\and
Laboratoire d'Astrophysique de Marseille,
38 rue F. Joliot-Curie, 13388 Marseille Cedex 13, France              
}

\authorrunning{M. Juvela et al.}

\date{Received September 15, 1996; accepted March 16, 1997}

\abstract
{
In the project Galactic Cold Cores we are carrying out Herschel
photometric observations of cold regions of the interstellar clouds as
previously identified with the Planck satellite. The aim of the
project is to derive the physical properties of the population of cold
clumps and to study its connection to ongoing and future star
formation.
}
{
We examine the cloud structure around the Planck detections in 71
fields observed with the Herschel SPIRE instrument by the summer of 2011.
We wish to determine the general physical characteristics of the
fields and to examine the morphology of the clouds where the cold high
column density clumps are found.
}
{
Using the Herschel SPIRE data, we derive colour temperature and column
density maps of the fields. Together with ancillary data, we examine
the infrared spectral energy distributions of the main clumps. The
clouds are categorised according to their large scale morphology. With
the help of recently released WISE satellite data, we look for signs
of enhanced mid-infrared scattering (`coreshine'), an indication of
growth of the dust grains, and have a first look at the star formation
activity associated with the cold clumps.
}
{
The mapped clouds have distances ranging from $\sim$100\,pc to several
kiloparsecs and cover a range of sizes and masses from cores of less than
10\,$M_{\sun}$ to clouds with masses in excess of 10000\,$M_{\sun}$.
Most fields contain some filamentary structures and in about half of
the cases a filament or a few filaments dominate the morphology. In
one case out of ten, the clouds show a cometary shape or have sharp
boundaries indicative of compression by an external force. The width
of the filaments is typically $\sim$ 0.2--0.3\,pc. However, there is
significant variation from 0.1\,pc to 1\,pc and the estimates are
sensitive to the methods used and the very definition of a filament. 
Enhanced mid-infrared scattering, coreshine, was detected in four
clouds with six additional tentative detections. The cloud LDN\,183 is
included in our sample and remains the best example of this
phenomenon. About half of the fields are associated with active star
formation as indicated by the presence of mid-infrared point sources.
The mid-infrared sources often coincide with structures whose
sub-millimetre spectra are still dominated by the cold dust.
}

\keywords{
ISM: clouds -- Infrared: ISM -- 
Submillimeter: ISM -- dust, extinction -- Stars: formation -- 
Stars: protostars
}

\maketitle
%

\section{Introduction}

The main phases of the star formation process are largely understood, starting from
molecular clouds and progressing via dense cores down to protostellar collapse \citep{McKee2007}. This
is thanks to the detailed observations of the nearest low mass and intermediate
mass star formation regions and, on the other hand, sophisticated numerical
modelling. However, the formation of each star is an individual process and, for
a global view, extensive surveys of different environments and of the
different phases of the star formation process are needed.

One major question is how the properties of star formation depend on the initial
conditions within the cold molecular clouds. Direct links should exist between the
physics of the prestellar clumps and the star formation efficiency, the mode of
star formation (clustered vs. isolated), and the masses of the born stars
\citep{Elmegreen2011, PadoanNordlund2011a, NguyenLuong2011}. The general time
scales of star formation have been extensively studied. Eventually, one has to
progress to the examination of the finer details, how the processes vary between
different environments and, in each case, what is the interplay between turbulence,
magnetic fields, kinematics, and gravity. Similarly, our understanding of the role
of external triggering is incomplete. How does the influence of supernovae and
stellar winds (and of galactic density waves, colliding flows of HI gas, etc.) vary
between different regions of the Milky Way and is this reflected in the timescales
and the stellar initial mass function?

The Planck satellite \citep{Tauber2010} has made a new approach to the
study of the earliest stages of star formation possible. By mapping the whole sky at several
sub-millimetre wavelengths, with high sensitivity and small beam size (below
5$\arcmin$ at the highest frequencies), Planck is providing data for a {\em global} census of the coldest
component of the interstellar medium. The selection of the cold ($T_{\rm
dust}<$14\,K) and `compact' (close to beam size) objects has led to a list of more
than 10000 objects. Because of the limited resolution, this Cold Clump Catalogue of
Planck Objects \citep[C3PO, see][]{PlanckI} is not dominated by cores
(pre-stellar or only starless cores at sub-parsec scales) but by $\sim$1\,pc sized
clumps and even larger cloud structures extending to even tens of pc in size.
However, the low temperatures of the objects (below 14\,K) are possible only for the denser, less evolved
regions that are well shielded from the interstellar radiation field. Therefore,
the objects detected by Planck are likely to contain one or several cores, many of
which will be pre-stellar or in the early stages of protostellar evolution. This Planck
survey constitutes the first unbiased census (in terms of sky coverage) of
possible future star forming sites and provides a good starting point for global
studies addressing the pre-stellar phase of cloud evolution.

Within the {\it Herschel} Open Time Key Programme {\em Galactic Cold Cores}, we are
mapping selected Planck C3PO objects with the {\it Herschel} PACS and SPIRE
instruments (100--500\,$\mu$m). The higher spatial resolution of {\it Herschel}
 \citep{Poglitsch2010, Griffin2010} makes it possible to examine the structure of
the sources which gave rise to the Planck detections, often resolving the individual cores. The
inclusion of shorter wavelengths (down to 100\,$\mu$m)  help to
determine the physical characteristics of the sources and their environment, and to
investigate the properties of the interstellar dust grains. Our {\it Herschel}
survey will eventually cover some 120 fields between 12$\arcmin$ and one degree in
size and will altogether cover approximately 350 individual Planck detections of cold
clumps.
Preliminary results of this follow-up have been presented in \citet{Juvela2010,
Juvela2011} (papers I and II, respectively) from the Herschel Science Demonstration Phase observations (three
fields there called PCC288, PCC249, and PCC550) and on a sample of ten Planck
sources in \cite{PlanckII}.

In this paper we present results for the first 71 fields that were observed with the SPIRE
instrument\footnote{SPIRE has been developed by a consortium of institutes led by Cardiff
University (UK) and including Univ. Lethbridge (Canada); NAOC (China); CEA, LAM (France);
IFSI, Univ. Padua (Italy); IAC (Spain); Stockholm Observatory (Sweden); Imperial College
London, RAL, UCL-MSSL, UKATC, Univ. Sussex (UK); and Caltech, JPL, NHSC, Univ. Colorado
(USA). This development has been supported by national funding agencies: CSA (Canada); NAOC
(China); CEA, CNES, CNRS (France); ASI (Italy); MCINN (Spain); SNSB (Sweden); STFC (UK);
and NASA (USA)} (250, 350, and 500\,$\mu$m) by the summer 2011. We concentrate on the large
scale structure of the clouds and the general characteristics of the main clumps. We
compare the sub-millimetre emission with other available far-infrared data, especially the
AKARI \citep{Murakami2007} satellite far-infrared maps, and use 
Wide-field Infrared Survey \citep[WISE][]{Wright2010} satellite observations (3.6--22\,$\mu$m) to
look for regions where enhanced mid-infrared scattering could indicate an increase in the
size of the dust grains. This so-called coreshine phenomenon was first detected with
Spitzer data of LDN183 \citep[see][]{Steinacker2010}, a cloud also included in the present
sample. The point sources detected in the mid-infrared WISE data also serve as an indicator
of the ongoing star formation.

The structure of the paper is the following. The observations are
described in Sect.~\ref{sect:obs}. The main results are presented in
Sect.~\ref{sect:results}, starting with the analysis of the large
scale properties of the clouds. These include the estimation of the
colour temperatures (Sect.~\ref{sect:temperature}), the column
densities, and cloud masses (Sect.~\ref{sect:colden}), all based on
the SPIRE observations. In Sect.~\ref{sect:categories} we describe a
categorisation of the target fields that is based on the main cloud
morphology. In Sect.~\ref{sect:infrared} we characterise the fields
further with the help of infrared data. We examine the properties of
the selected clumps, present their spectral energy distributions and
masses (Sect.~\ref{sect:clumps}) and then look for coreshine in the
mid-infrared data (Sect.~\ref{sect:coreshine}).  The results
concerning the general nature of the fields and the connection to the star
formation are discussed in Sect.~\ref{sect:discussion}. The final
conclusions are presented in Sect.~\ref{sect:conclusions}. The online
appendices contain additional figures, including the surface
brightness and the column density maps and the SEDs of selected
clumps.

A detailed analysis of all compact sub-millimetre clumps (including the core mass
spectra) and  of the young stellar objects associated with the
cold clumps will be presented in future papers. An in-depth study of dust emission
properties (opacity and emissivity spectral indices) will also be presented
later, together with the data from observations with the {\it Herschel} PACS
instrument.

\section{Observations}  \label{sect:obs}

\subsection{Target selection}

The Planck satellite is performing surveys of the full sky at nine wavelengths
between 350\,$\mu$m and 1\,cm \citep{Tauber2010}. The analysis of the first two
sky surveys led to the detection of more than 10000 compact sources of cold dust
emission as described in \citet{PlanckI}. The detections are based on the
sub-millimetre cold dust signature that becomes visible when the warm extended
emission is subtracted using the IRAS 100\,$\mu$m maps as its spatial template
and the average spectrum of the region as its spectral template
\citep{Montier2010}. The detection procedure limits the maximum size of the
detected clumps to $\sim$12$\arcmin$. Together with the IRAS 100\,$\mu$m data,
the Planck measurements have been used to estimate the colour temperatures of the
sources. The source distances are estimated using several methods 
\citep[see][for details]{PlanckI} that include association with known
molecular clouds complexes or individual sources and the reddening of background
stars at optical or near-infrared wavelengths \citep{Marshall2009,
McGehee2012}.
Altogether distance estimates  exist for approximately one third of the C3PO catalogue.

The pre-selection of targets for the {\it Herschel} observations was based on a
binning of Planck cold clumps with the following parameter boundaries: $l$= 0, 60,
120, and 180 degrees, $|b|=$1, 5, 10, and 90 degrees, $T_{\rm dust}$=6, 9, 11, and
14\,K, and $M=$0, 0.01, 2.0, 500, 10$^6$\,$M_{\sun}$. Here $T_{\rm dust}$ is the
clump temperature obtained after the subtraction of the warm emission component
\citep[see][]{PlanckI}. The binning was used to ensure a full coverage of the
respective parameter ranges while at the same time weighting the sampling towards
sources at high latitudes and with extreme values of the mass. The lowest mass bin
($M$=0) was reserved for sources without distance estimates and, therefore, without
any available mass estimates. The Galactic latitudes $|b|<1^{\degr}$ were excluded
because those regions will be covered by the {\it Herschel} key programme Hi-GAL
\citep{Molinari2010}. Similarly, the regions covered by the other key programmes
like the Gould Belt survey \citep{Andre2010} or HOBYS \citep{Motte2010} were avoided.

\begin{figure*}
\centering
\includegraphics[width=17cm]{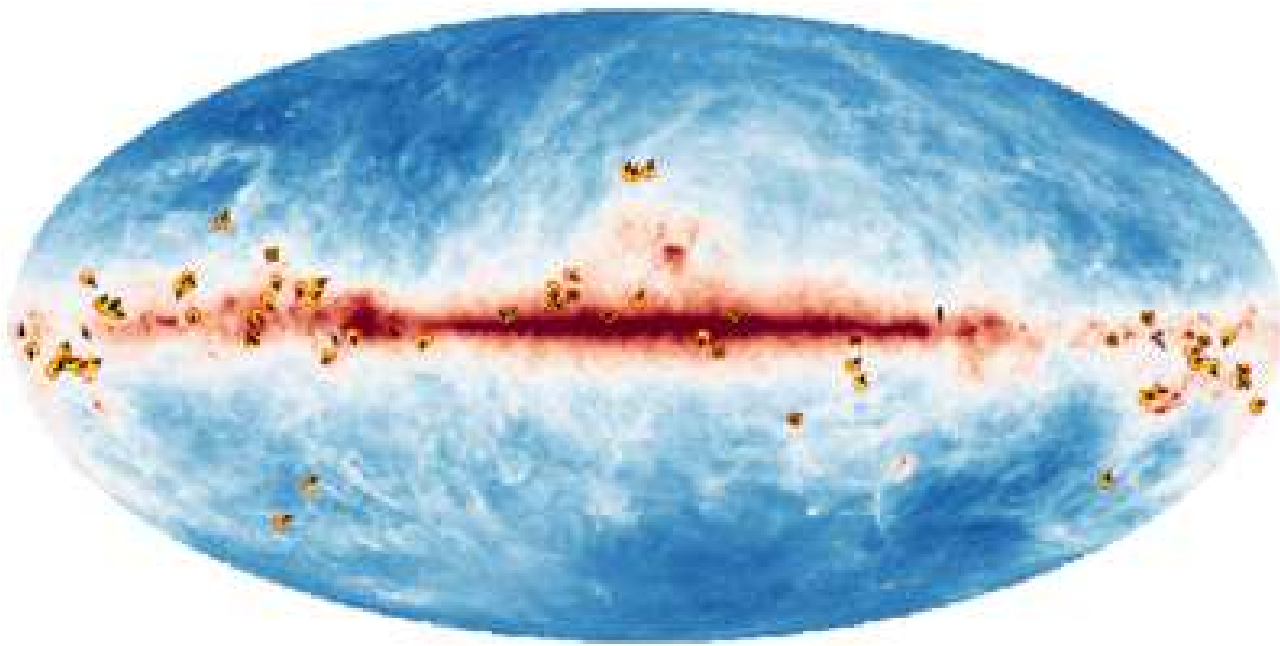}
\caption{
The locations of the observed fields. The background is the IRAS
100\,$\mu$m map and the circles denote the positions of the fields. The
arrows drawn on each circle indicate the distances, starting with 0\,pc
for the upright direction, one clockwise rotation corresponding to 2\,kpc.
The six sources with distances larger than 1.5\,kpc are drawn with smaller
symbols. The sources without reliable distance estimates are marked with
crosses.
}
\label{fig:allsky}%
\end{figure*}

\begin{figure}
\centering
\includegraphics[width=7.5cm]{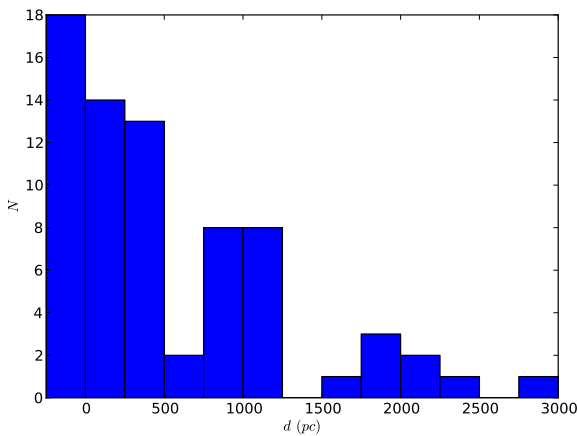}
\caption{
The distance distribution of the fields. The bin at negative values
corresponds to the sources without a distance estimate.
}
\label{fig:distances}%
\end{figure}

The above procedure results in 108 bins, from which one target per bin was
selected for {\it Herschel} follow up observations. Within each bin, the sources were inspected
visually, comparing the Planck maps with IRAS and AKARI dust emission maps,
extinction maps calculated using 2MASS stars, and CO emission maps
\citep{Dame2001, Fukui1999}, when available. These data were used to confirm the
reliability of the original detection.
Some preference was given to fields containing several Planck detected clumps.
The final selection of 108 fields covers some 350 cold clumps from the C3PO
catalogue. 
Out of the full list of 108 fields, in this paper we use the
SPIRE observations of the 71 fields for which the observations were
completed by the {\it Herschel} observational day 721. In this
sample, distance estimates exist for 75\% of the fields (53 out of
71), partly because a literature search has resulted in a few
additional distance estimates.
Figure~\ref{fig:allsky} shows the positions of the sources on the sky
and Fig.~\ref{fig:distances} the distribution of their distances. For
the fields with distance estimates, the median value is 450\,pc. 
When distances were estimated from the stellar reddening
\citep{Marshall2009, McGehee2012}, every effort was made to identify
the extinction features most likely to be associated with the main
clouds. However, the fields may contain clouds also at other
distances, especially at the low Galactic latitudes. 
The distance distribution of all Planck clumps is smooth over the
shown range of distances \citep{PlanckI}. The small number statistics
and the uncertainties of the distance estimates (e.g., the use of
different methods in different distance ranges) may contribute to the
dips seen in Fig.~\ref{fig:distances} at $\sim$600\,pc and
$\sim$1400\,pc. The distribution is also affected by the exclusion of
the areas covered by the other Herschel key programmes. In particular,
the avoidance of the Gould Belt clouds reduces the number of objects
close to 500\,pc.

\subsection{{\it Herschel} observations}

In this paper we use data from the first 71 fields observed with the
{\it Herschel} SPIRE instrument. The observations consist of 250, 350,
and 500\,$\mu$m maps with the map sizes ranging from 37 to 77 arc
minutes. The observations of the first three
sources were done as part of the Science Demonstration Phase in
November and December 2009 and the latest were performed at the
beginning of May 2011. The fields are listed in
Table~\ref{table:main}.
The {\it Herschel} observations were reduced with the {\it Herschel} Interactive Processing
Environment HIPE v.7.0. using the official pipeline\footnote{HIPE is a joint development by
the Herschel Science Ground Segment Consortium, consisting of ESA, the NASA Herschel
Science Center, and the HIFI, PACS and SPIRE consortia.}.
The resulting SPIRE maps are the product of direct projection onto the sky and
averaging of the time ordered data (the HIPE naive map making routine)
\footnote{The reduced data will become available through the ESA web
site {\em http://herschel.esac.esa.int/UserReducedData.shtml}. Further
information is available on the home page of the Cold Cores
project {\em
https://wiki.helsinki.fi/display/PlanckHerschel/The+Cold+Cores}.}
In order of wavelength, the resolution of the SPIRE maps is 18$\arcsec$,
25$\arcsec$, and 37$\arcsec$, respectively.

We rely on the {\it Herschel} calibration of the data. 
The accuracy of the absolute calibration of the SPIRE observations is
expected to be better than 7\%\footnote{SPIRE Observer's manual, \\
{\em http://herschel.esac.esa.int/Documentation.shtml}.}
However, in order to determine the absolute zero point of the
intensity scale, we carried out a comparison with Planck satellite
observations that were complemented with the IRIS version of the IRAS
100\,$\mu$m data \citep{MAMD2005}. 
The Planck and IRIS measurements were interpolated to the {\it Herschel} wavelengths
using fitted modified blackbody curves, $B_{\nu}(T_{\rm dust}) \nu^{\beta}$, with a fixed
value of the spectral index, $\beta=2.0$. The linear correlations between
{\it Herschel} and the reference data were extrapolated to zero Planck (+IRIS) surface
brightness to determine the offsets for the {\it Herschel} maps. The uncertainties of
the offsets were obtained from the formal errors of these fits. The errors are
typically $\sim$1\,MJy\,sr$^{-1}$ at 250\,$\mu$m and, corresponding to the
decreasing surface brightness values at the longer wavelengths. 
The derived intensity zero point is independent of the Planck calibration and any
multiplicative errors included in the comparison. Similarly, the offset determination
depends very little on the assumed value of $\beta$. Colour corrections can have an
effect on the quality of the linear correlation between Planck and {\it
Herschel} data. The data were first colour corrected assuming a constant colour
temperature of 15\,K. The offsets were later re-evaluated using the actual dust
colour temperature estimates (Sect.~\ref{sect:temperature}).

\subsection{Other infrared data}

In the infrared range, we use data from the AKARI and WISE satellites.

From the AKARI survey \citep{Murakami2007} we use observations from the all-sky
survey made with the FIS instrument. These include observations in the narrow
band filters N60 and N160 (central wavelengths 65\,$\mu$m and 160\,$\mu$m) and in
the wide band filters WideS and WideL (central wavelengths 90\,$\mu$m and
140\,$\mu$m). The spatial resolution of the data ranges from 37$\arcsec$ at
65\,$\mu$m to 61$\arcsec$ at 160\,$\mu$m. The accuracy of the calibration is
assumed to be $\sim$30\%. AKARI data are available for all the fields.

The WISE satellite \citep{Wright2010} has four bands centred at 3.4, 4.6, 12.0, and
22.0\,$\mu$m with spatial resolution ranging from 6.1$\arcsec$ at the shortest wavelength
to 12$\arcsec$ at 22\,$\mu$m. The first public release of WISE satellite data took
place in April 2011 and it provides data for 55 of our 71 fields.  We use the WISE
3.4\,$\mu$m, and 4.6\,$\mu$m data to look for signs of enhanced mid-infrared
scattering and the WISE 12.0\,$\mu$m and 22.0\,$\mu$m data to characterise the
mid-infrared dust emission and to look for indications of ongoing star formation. 
The data were converted to surface brightness units with the conversion factors
given in the explanatory supplement \citep{Cutri2011}. The calibration uncertainty
is $\sim$6\% for the 22\,$\mu$m band and less for the shorter wavelengths.

We calculated dust extinction maps for each field with stars from the
Two Micron All Sky Survey \cite[2MASS,][]{Skrutskie2006}, using the
NICER method \citep{Lombardi2001}. The values of $A_{\rm V}$ were
obtained from near-infrared (NIR) colour excess measurements assuming an extinction
law with $R_{\rm V}$=3.1. With the assumption of $R_{\rm V}$=5.5,
possibly more appropriate for the densest regions, the $A_{\rm V}$
estimates would increase by $\sim$16\% \cite[see][]{Juvela2011}.
The spatial resolution of all the produced extinction maps is two arc
minutes. For distant sources, the extinction of the target clouds
cannot be reliably reproduced because of the poor resolution and the
increasing number of foreground stars. The same applies to some extent
to the nearby, high latitude targets. No special steps have been taken
to eliminate the contamination by foreground stars \citep[see,
e.g.,][]{Schneider2011}, apart from the sigma clipping procedure
(clipping performed at the 3-$\sigma$ level)
included in the NICER method. Therefore, the extinction maps should
not be trusted unreservedly as tracers of the column density of the
examined clouds. On the contrary, the possible discrepancy between the
$A_{\rm V}$ map and the sub-millimetre emission serves as an indicator
of a large source distance.

\section{Results} \label{sect:results}

\subsection{Cloud properties derived from SPIRE}
\label{sect:results_SPIRE}

In this section we present the results derived from the SPIRE data. These include
colour temperature and column density maps and, when distance estimates are
available, estimates of the cloud masses. In Sect.~\ref{sect:filaments}, we
examine the properties of the filamentary cloud structures.

\subsubsection{Colour temperature maps} \label{sect:temperature}

The colour temperature maps of the large grain emission are estimated using
the 250\,$\mu$m, 350\,$\mu$m, and 500\,$\mu$m SPIRE maps with the main goal
of identifying the coldest clumps. The data cover only the longer wavelength
side of the dust emission peak which means that the derived temperatures may
not be very accurate, especially for warm regions with temperatures close to
20\,K or above. On the other hand, the absence of shorter wavelengths
decreases the bias that results from temperature variations along the line of
sight \citep{Shetty2009a, Shetty2009b, Malinen2011}. The statistical noise
will be larger than, for example, if one included the PACS bands, but the
values will be more representative of the bulk of the cold dust. The absence
of data below or near 100$\mu$m also avoids the problem of a possible
contribution from stochastically heated small grains.

The maps were convolved to a 40 arcsec resolution and, for each pixel, the
SED was fitted with $B_{\nu}(T_{\rm dust}) \nu^{\beta}$ keeping the spectral
index $\beta$ at a fixed value of 2.0. Several studies have suggested that
the spectral index may increase in cold and dense environments 
\citep[e.g.][]{Dupac2003, Desert2008, PlanckI}. If this
anticorrelation between the dust temperature and the spectral index is true
also in our fields, the derived temperature maps will underestimate the range
of temperature variations and, in particular, will overestimate the
temperature of the coldest regions, leading to an underestimation of the
masses of these regions \citep{PlanckI}.
However, the temperature maps will still fulfill their main purpose,
identifying the major relative temperature variations within the regions. 

\begin{figure}
\centering
\includegraphics[width=7.7cm]{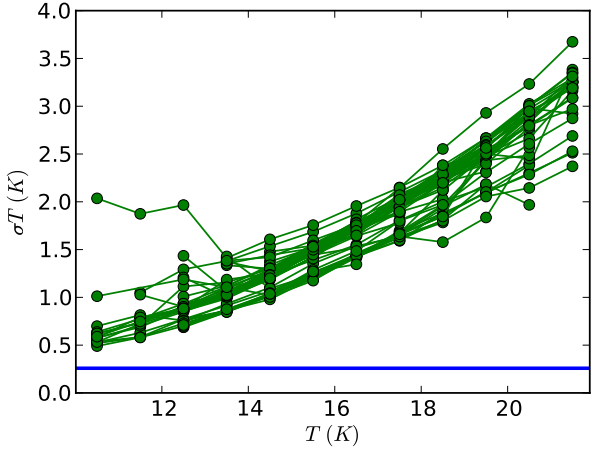}
\caption{
Uncertainty of the colour temperatures estimated with a Monte Carlo study.
The curves show, for each field, the mean error of the temperature if the
uncertainty of the surface brightness data were 13\%. The horizontal line
is the 1-$\sigma$ value (calculated over all fields) of the bias
associated with the uncertainty of the intensity scale zero points.
}
\label{fig:temperature_errors}
\end{figure}

The effect of statistical errors was examined with the help of Monte-Carlo
simulations where a 13\% uncertainty in the surface brightness values was
assumed. Based on the correlations between the different bands, this is a
conservative estimate and probably even twice the value of the typical true
noise. The 13\% uncertainty of the surface brightness values would
translate to an error below 1\,K in cold regions, the error increasing to
$\sim$3\,K at 20\,K (Fig.~\ref{fig:temperature_errors}).
The calibration accuracy of the SPIRE data is estimated to be 7\%. This is
well within the above limits.

The effect of the intensity zero points was estimated in a similar fashion, examining
realisations of the temperature maps when the zero points were modified in
accordance with the uncertainty of the comparison of {\it Herschel} and
Planck/IRIS data. This uncertainty was less significant, typically less than
0.5\,K and $\sim$1\,K in a few cases (G130.42-47.07, G161.55-9.30,
G37.49+3.03). However, the effects on the relative temperature variations
within a given field are, of course, much smaller. The temperatures can be
further affected by mapping artifacts, e.g., residual striping. The striping
is usually visible only towards the map edges where its effect can rise to
$\sim$0.5\,K.

Two examples of calculated temperature maps are seen in frame $a$
of
Figs.~\ref{fig:ids_1} and \ref{fig:ids_1b}. The colour temperature
maps of all the fields are presented in Appendix~\ref{sect:appendix_ids}
(online edition).

\begin{figure*}
\centering
\includegraphics[width=16cm]{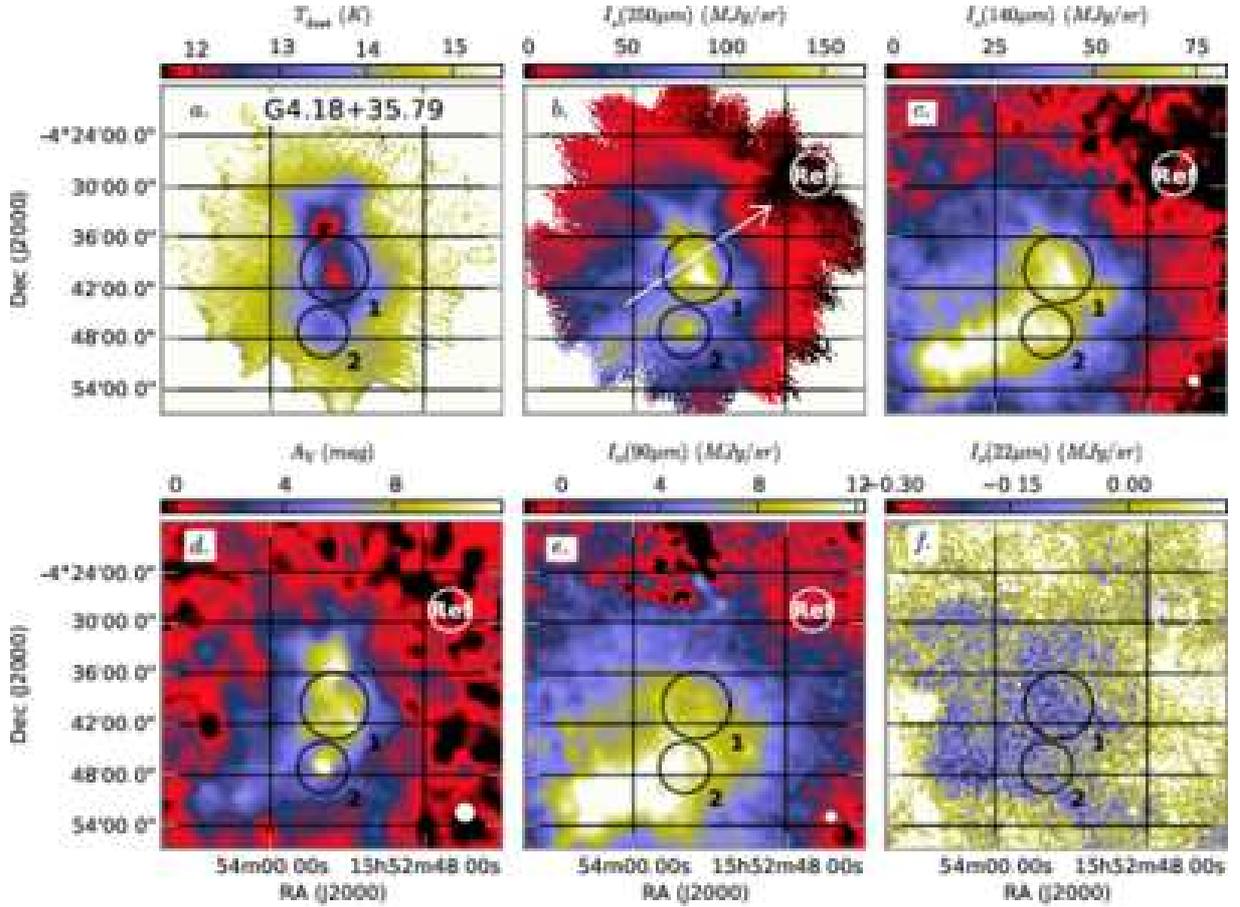}
\caption{
Selected observed and derived quantities for the field G4.18+35.79
(LDN~134). Shown are the colour temperature map derived from SPIRE data
(frame $a$, 40$\arcsec$ resolution), the 250\,$\mu$m SPIRE surface
brightness map (frame $b$, 18$\arcsec$ resolution), the AKARI wide filter
maps at 140\,$\mu$m and 90\,$\mu$m (frames $c$ and $e$), the visual
extinction $A_{\rm V}$ derived from 2MASS catalog stars (frame $d$), and
the WISE 22\,$\mu$m intensity (frame $f$).
For the other fields, see Appendix~\ref{sect:appendix_ids}. 
The respective beam sizes are indicated in the lower right hand corner of
each frame. The positions of selected clumps (black circles; see
Sect.~\ref{sect:clumps}) and the reference regions used for background
subtraction (white circle; see Sect.~\ref{sect:clumps}) are also shown. 
In frame $b$, the arrow indicates the position of the stripe plotted in
Fig.~\ref{fig:stripes_1}.
}
\label{fig:ids_1}%
\end{figure*}

\begin{figure*}
\centering
\includegraphics[width=16cm]{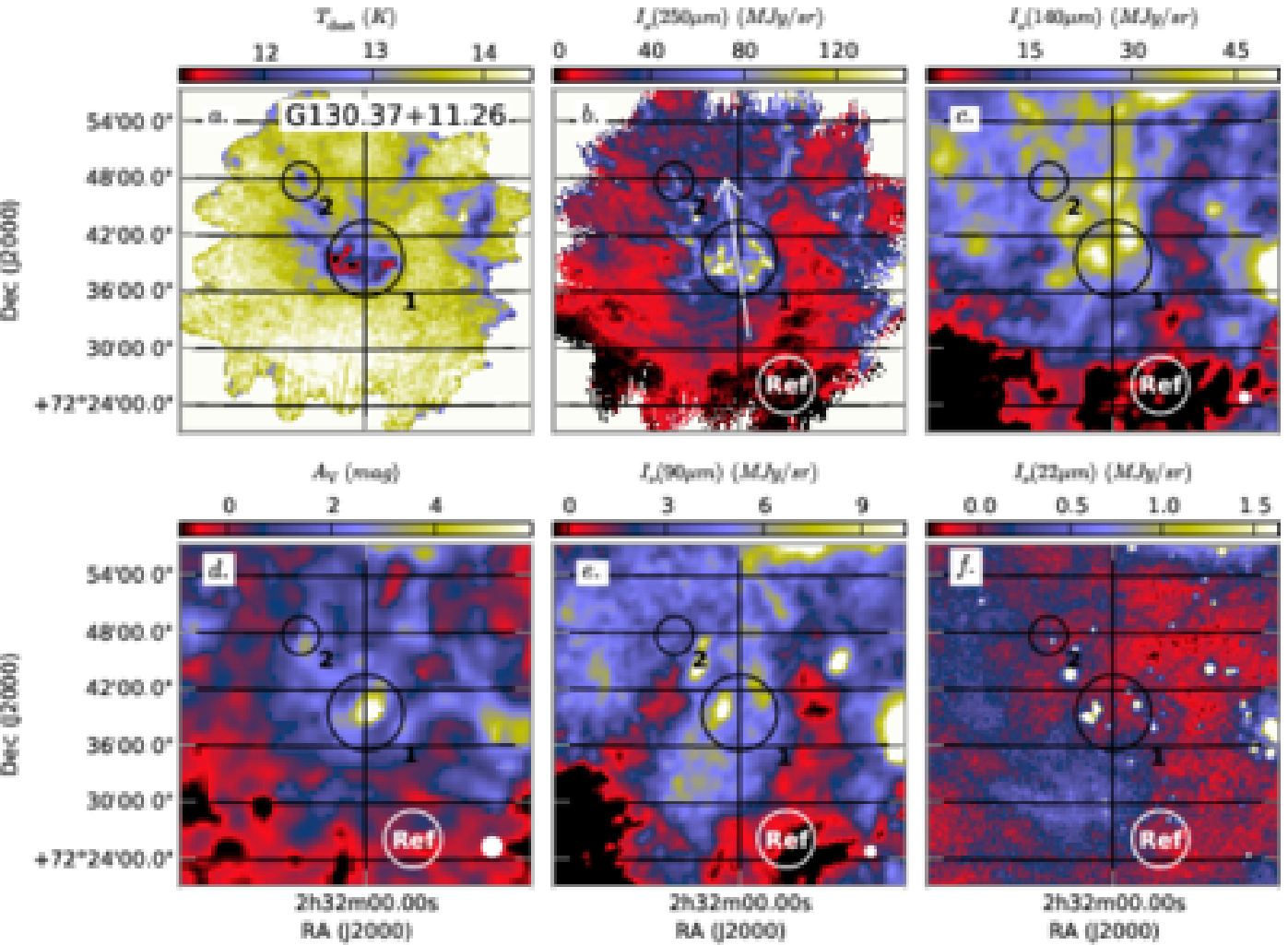}
\caption{
Selected observed and derived quantities for the field G130.37+11.26 (see
Fig.~\ref{fig:ids_1} for details).
}
\label{fig:ids_1b}%
\end{figure*}

\subsubsection{Column density maps and masses} \label{sect:colden}

The column densities averaged over a 40$\arcsec$ beam are calculated using the formula
\begin{equation}
    N({\rm H}_2) = \frac{ I_{\nu} }{ B_{\nu}(T) \kappa \mu m_{\rm H}},
\end{equation}
with the intensity and temperature of the previous SED fits
(Sect.~\ref{sect:temperature}) and using $\mu=2.33$ for the particle mass per hydrogen molecule.
We use a dust opacity $\kappa$ obtained from the formula
0.1\,cm$^2$/g\,($\nu$/1000\,GHz)$^{\beta}$ that is applicable to high density
environments \citep{Beckwith1990}. 
The value of dust opacity is uncertain and is observed to vary
from region to region \citep[e.g.][]{Kramer2003, Stepnik2003,
Lehtinen2004, Lehtinen2007, Ridderstad2006, Martin2011}. In diffuse
medium the expected value of $\kappa$ is lower by a factor of two
\citep{Boulanger1996}. However, the value chosen here is close to the
predictions and the observations of dense clouds
\cite[e.g.][]{Ossenkopf1994, Nutter2006, Nutter2008} and also is the
value used in \cite{PlanckI, PlanckII}. 
The column density maps of the first eleven
sources are shown in Fig.~\ref{fig:colden_1} and for the rest in
Appendix~\ref{sect:appendix_colden} (online edition). The resolution of the maps is 40$\arcsec$.

Mass estimates are possible for the sources with distance estimates, i.e., 59 out
of the 71 fields (see Table~\ref{table:main}). Masses are calculated for the
entire field, for some major filaments (see Sect.~\ref{sect:filaments}), and for
one or two representative clumps within each field (see Sect.~\ref{sect:clumps}).
The mass estimates are listed in Table~\ref{table:mass}. These mass estimates are
intended only for general characterization of the target fields,  while more detailed
analysis of all clumps, including their mass spectra, will be presented in a
later publication.

\begin{figure*}
\centering
\includegraphics[width=14cm]{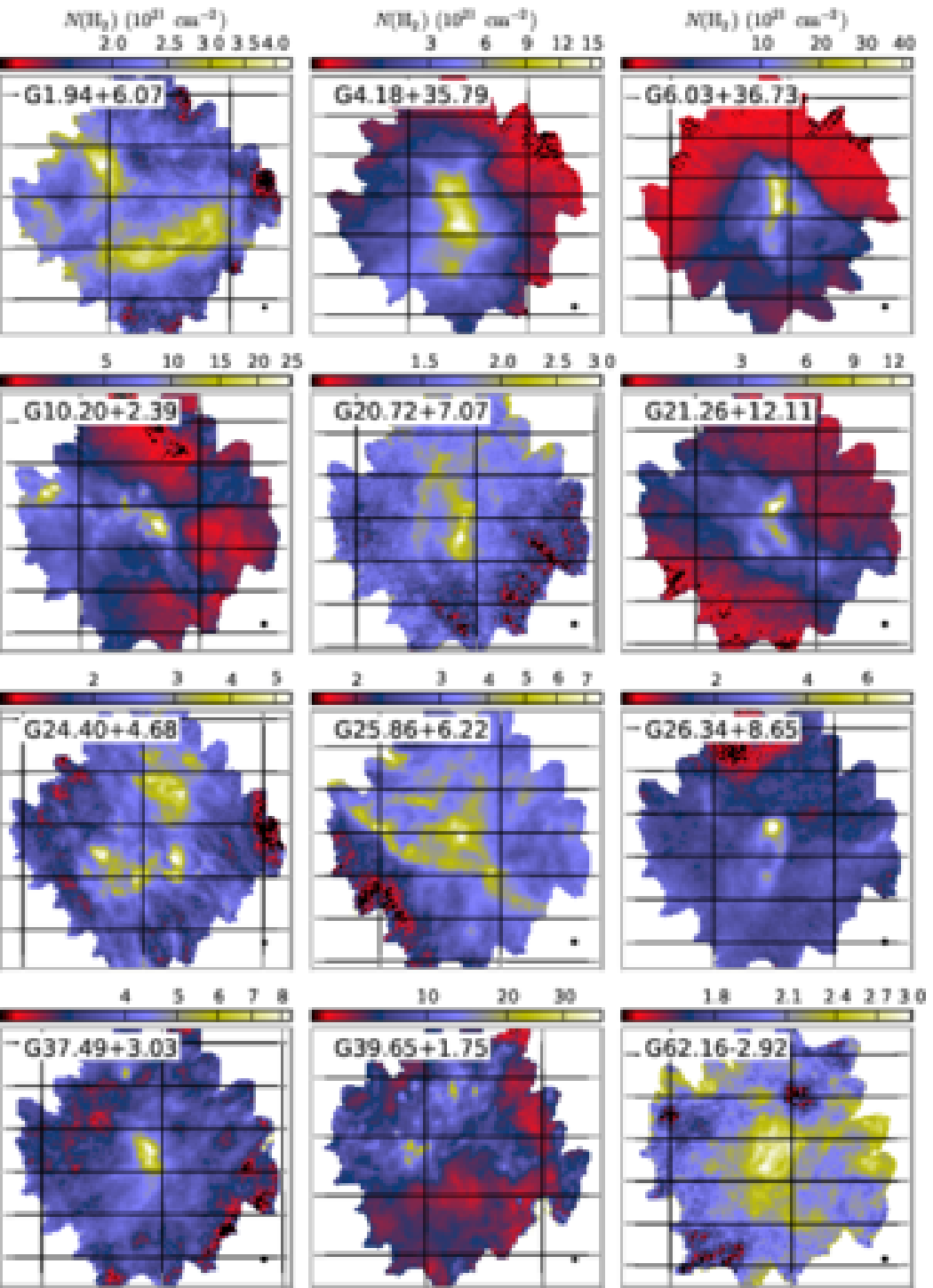}
\caption{
Column density maps of the 12 first fields in the order of increasing
Galactic longitude. The map resolution is 40$\arcsec$ and the areas are
the same as in Fig.~\ref{fig:ids_1} and
Appendix~\ref{sect:appendix_ids}  (online edition).
The colour scale has a square root scale.
The column densities of the other fields are shown in
Appendix~\ref{sect:appendix_colden}.
}
\label{fig:colden_1}%
\end{figure*}

\subsubsection{Categorisation of the source fields} \label{sect:categories}

The sources, as seen through the sub-millimetre emission, exhibit different
morphologies from isolated unstructured clumps to more complex configurations. We
grouped the sources to non exclusive sets of cometary, filamentary, isolated, and
complex sources (tags $C$, $F$, $I$, and $X$ in Table~\ref{table:main}). 

The category of cometary sources refers to the shape of the major clump or clumps
that suggests deformation by an external pressure that has resulted in a shape
typical of cometary globules \citep{Hawarden1976} (e.g. LDN~1780 in
Fig.~\ref{fig:ids_68}, online edition). We distinguish sources that show  more
local indication of direct dynamic interaction, e.g., in the form of a sharp cloud
boundary, without the overall cloud shape being cometary (tag $B$; e.g. field
G25.86+6.22 shown in Fig.~\ref{fig:ids_06},  online edition). About one field in ten shows clear
signs of either type of a dynamical interaction.

In the group marked as filamentary (tag $F$; e.g. field G82.65-2.00 in
Fig.~\ref{fig:ids_13}  in online edition), the clump or clumps are located inside narrow, elongated
structures or the filaments themselves are the dominant source of dust emission. The
filaments will be discussed in more detail below but one must already note that the
definition of a filamentary cloud is not clear and, as a result of turbulent
motions, some degree of filamentary structure is visible in all fields. In about
half of the fields the overall morphology is dominated by a single or a couple of
filaments.

We denote fields as complex when they contain several clumps that are not clearly
arranged onto a small number of filaments (tag $X$, e.g. field G86.97-4.06 in
Fig.~\ref{fig:ids_14} in online edition). At low latitudes, this also could be caused by line-of-sight
confusion without it being an intrinsic property of the clouds observed. On the
other hand, the isolated sources (tag $I$) are typically found at  intermediate
and high latitudes.

Few fields can be assigned unambiguously to a single category and in most cases
one can identify characteristics of more than one category as indicated in the
eighth column of Table~\ref{table:main}. The classification is, of course, a
subjective one and is made only to help the discussion below.

\subsubsection{Properties of the filaments} \label{sect:filaments}

Half of the examined fields are categorised as having a prominently
filamentary structure (see Sect.~\ref{sect:categories} and
Table~\ref{table:main}). We examine here the properties of selected
cloud segments. The selection is carried out using column density maps
that are obtained by combining, without further convolution, the
250\,$\mu$m surface brightness data (resolution $\sim 18\arcsec$) and
the colour temperature estimates (resolution 40$\arcsec$, see
Sect.~\ref{sect:temperature}). This gives a nominal resolution of
$\sim 20\arcsec$ although, of course, the column density estimates
depend on the temperature information that is available only at half
of this resolution. As the dust temperature is expected to decrease
towards the centre of the filaments, the low resolution of the
temperature data is likely to cause the central column densities to be
underestimated and the column density profiles to appear flatter than
in reality. However, the colour temperatures are known to be biased
estimates of the mass weighted average dust temperature
\citep{Shetty2009a, Shetty2009b, Malinen2011} and the low resolution
of the temperature data may not be the main source of uncertainty. We
will return to this question in Sect.~\ref{sect:discussion}.

We select from each of the maps the most prominent filament (or, more
generally, elongated structure) and trace by eye the path along its
central ridge. The column density profiles perpendicular to the path
are extracted at 20$\arcsec$ intervals. We examine the average column
density profile as well as its variations along the length of the
filament. The estimated width of the filaments depends on how far from
the centre of the filament the profile is followed and how the
background is subtracted. If the area is fixed in angular units, one
will probe larger linear scales for the more distant sources and,
because of the hierarchical structure of the ISM, the result will be a
correlation between the distance and the width estimate. To avoid
this, we use a fixed range of [-0.4\,pc,+0.4\,pc] around the peak of
the column density profile. This scale is adequate for most fields as 
it is resolved for the most distant sources while
remaining smaller than the map size in the case of the more nearby
targets.
The FWHM is measured after the subtraction of a constant background
level that is estimated as the minimum value within the
[-0.4\,pc,+0.4\,pc] interval.

We also fit the filament profiles with Plummer-like density profiles
\begin{equation}
\rho_{p}(r) = \frac{\rho_{c}}{[|1+(r/R_{\rm flat})^2|^{p/2}]}
\Rightarrow
N(r)  = A_{p} \frac{\rho_{c}R_{\rm flat}}{[1+(r/R_{\rm flat})^2]^{(p-1)/2}}
\end{equation}
\citep[see][]{Nutter2008, Arzoumanian2011}. In the equation $\rho_{\rm
c}$ is the central density, $R_{\rm flat}$ is the size of the inner
flat part, and $p$ describes the steepness of the profile.  For an
isothermal cylinder in hydrostatic equilibrium, the value of $p$
should be 4
\citep{Ostriker1964} but the values derived from observations are usually
 smaller \citep[e.g.][]{Arzoumanian2011}. The factor $A_{p}$ is
obtained from the formula
$A_{p} = \frac{1}{cos\, i} \int_{-\infty}^{\infty} \frac{du}{(1+u^2)^{p/2}}$,
which depends on the unknown inclination of the filament for which we
will assume a value of $i=0$. \cite{Arzoumanian2011} noted that while
this assumption does not affect the analysis of the shapes of the
profiles, the observed column density is on the average $\sim$60\%
higher than the column density measured perpendicular to the filament.
This could affect the estimates of the stability of the filaments. On
the other hand, it is clear that selection effects here favour small
inclination angles. The data are fitted with the Plummer profile
together with a linear background, keeping the centre position along
the filament profile as a free parameter. 
Because a linear background is part of the fit, these results do not
depend on the absolute level of the background but are still affected
by the extent of the fitted region, [-0.4\,pc, +0.4\,pc].

When we calculate the average profile of the whole filament segment,
the individual profiles are first aligned so that the peak column
density appears at the centre of the profile. This decision tends to
minimise the width of the average profile. 

\begin{figure*}
\centering
\includegraphics[width=16.5cm]{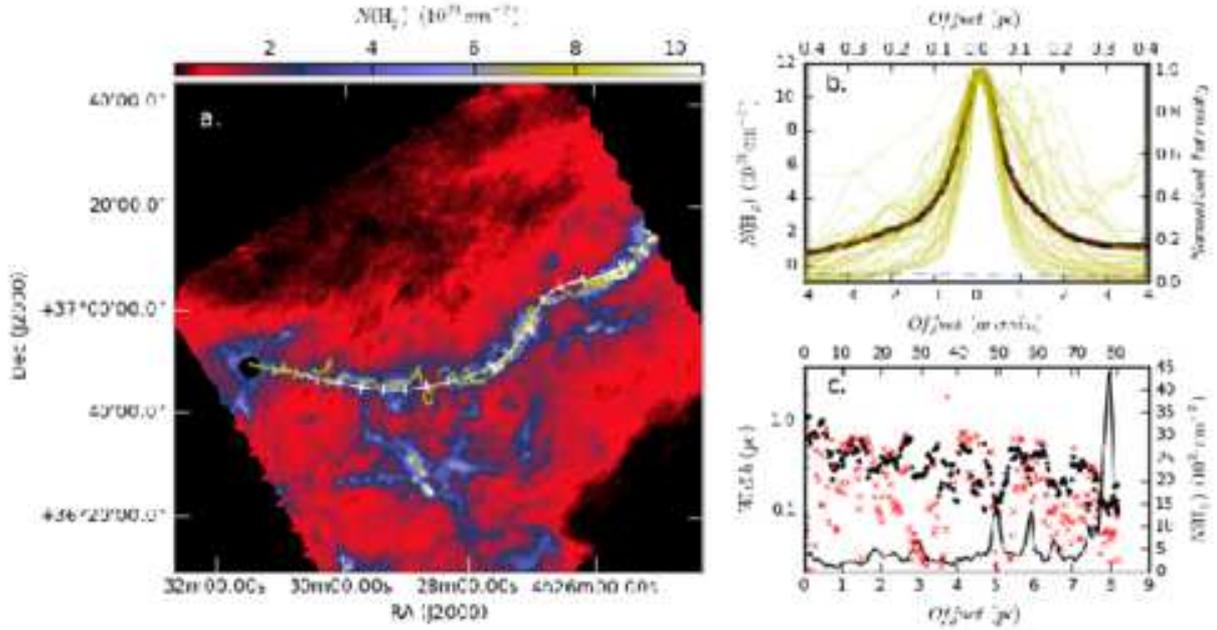}
\caption{
Properties of a filament in the field G163.82-8.44. The frame $a$ shows the
column density map. The white line shows the filament that was originally traced by eye
 and the yellow line follows the ridge that is formed by the peaks of
the column density profiles in the perpendicular direction. The black
filled circle indicates the start of the examined filament section. The
tick marks are drawn at 5 arcmin intervals. 
The frame $b$ shows the average column density profile of the filament
(black line), as well as the Plummer profile (the red dashed line on
top of the black line) that was fitted together with a linear baseline
(the blue dashed line) over the range -0.4\,pc to +0.4\,pc. The
yellow lines show individual column density profiles for 20\% of the cuts
with the highest column densities (values normalized to a peak value of one,
the right hand scale). The frame $c$ shows the FWHM values (the black
circles) and the parameter $R_{\rm flat}$ of the Plummer fit (the
red crosses), and the column density along the ridge of the filament (the
solid line and the right hand scale) as a function of the distance along
the filament. With the estimated distance of 350\,pc, the
SPIRE 250\,$\mu$m data resolution of 18$\arcsec$ corresponds to
0.03\,pc.
The plots for the other fields are shown in
Appendix~\ref{sect:appendix_filaments}.
}
\label{fig:filaments_1}%
\end{figure*}

Fig.~\ref{fig:filaments_1} shows  an example of the results for the
field G163.82-8.44. The frame $a$ shows,
on the column density map, the initial path that was selected by eye
(the white line). The beginning of the path is marked with a filled
circle and the end with an open circle. The yellow line traces the
peaks of the individual column density profiles. 

Figure~\ref{fig:filaments_1}$b$ shows the mean column density profile
and the fitted Plummer profile. The profile was obtained by averaging
the individual profiles where the column density rises above the
median value over all the profiles that were sampled at 20$\arcsec$
intervals. We also ignore those profiles where the column density peak
has shifted more than 5$\arcmin$ from the initial filament path. In
the same frame are drawn the individual profiles where the peak is
amongst the 20\% of the highest column densities along the filament
(the yellow lines). 

The frame $c$ shows the filament FWHM and the parameter $R_{\rm flat}$
of the Plummer profile fit as a function of the filament length. For
comparison, the column density along the filament is shown as a solid
line. Both the frames $b$ and $c$ indicate variations in the width of
the filament. These often are anticorrelated with the column density
(and, consequently, temperature) and are discussed further in
Sect.~\ref{sect:discussion_filaments}.

\begin{figure*}
\centering
\includegraphics[width=16.5cm]{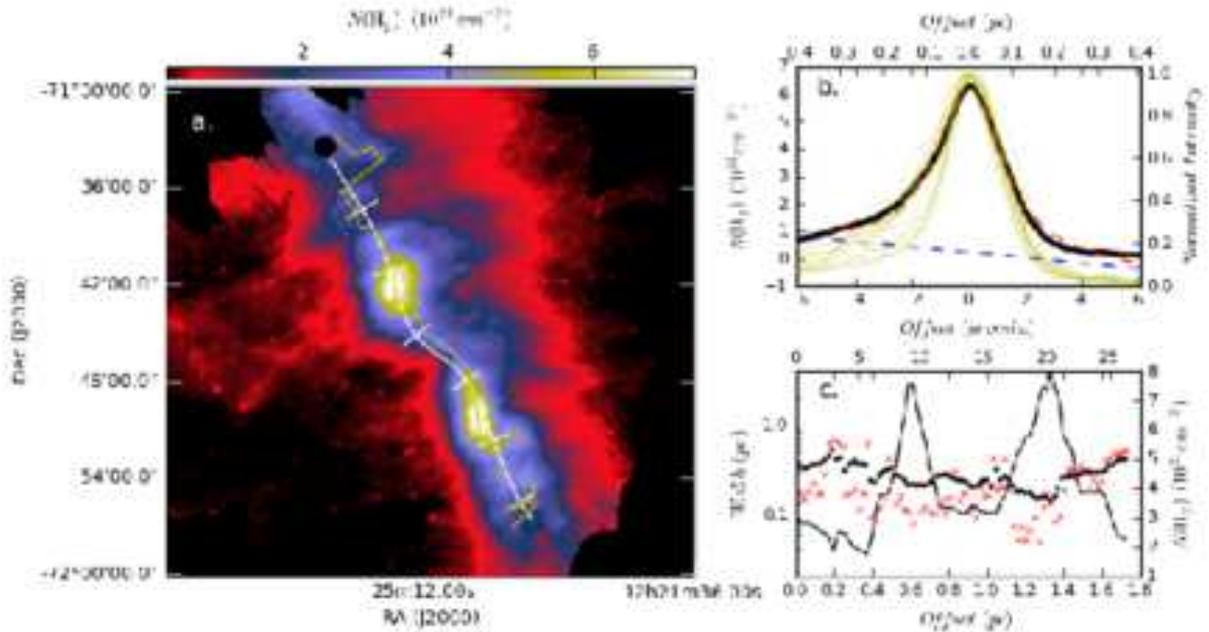}
\caption{
The filament in the field G300.86-9.00 (PCC550) that is part of
the Musca molecular cloud at the distance of 230\,pc. See
Fig.~\ref{fig:filaments_1} for the details.
}
\label{fig:filaments_2}%
\end{figure*}

Figure~\ref{fig:filaments_2} shows the corresponding results for the
more nearby field of G300.86-9.00 (PCC550). For the other fields the figures are
included in the Appendix~\ref{sect:appendix_filaments}. 

Table~\ref{table:filaments} lists the properties derived for the
average filament profile of each field where distance estimates were
available. These include the average column density (corresponding to
the average profile in Fig.~\ref{fig:filaments_1}b), the FWHM values
determined from data within 0.4\,pc of the filament centre, and the
parameters $R_{\rm flat}$ and $p$ from the fits of the Plummer
profile.
The figures show that there is often considerable variation in the
width along the length of the selected structures. Therefore, the
listed widths are not always representative of the cleanest filament
segments which tend to be more narrow.
The table includes further estimates of the mass per linear distance
that can be compared to the critical value of a gravitationally stable
filament, $M^{\rm crit}_{\rm line} = 2 c_{\rm s}^2/G$
\citep{Ostriker1964}. Instead of using the estimated dust temperature,
we use a sound speed $c_{\rm s}$ for a constant gas temperature of
12\,K that gives $M^{\rm crit}_{\rm line}=20$\,M$_{\sun}$\,pc$^{-1}$.
This is justified because, for most of the filament volume, the gas
and dust temperatures remain uncoupled \citep{Goldsmith2001,
JuvelaYsard2011a}. Note that this $M^{\rm crit}_{\rm line}$ is
applicable only in an isothermal case, in the absence of other
supporting forces like magnetic fields and turbulence. Of the 30
filaments listed in the table, 17 appear to be supercritical.

\subsection{Comparison with other infrared data}  \label{sect:infrared}

We have compared the sub-millimetre observations with the mid-infrared
and far-infrared data from the WISE and AKARI satellites. The shorter
wavelengths are more sensitive to temperature differences in the thin
surface layers of the clouds and can trace variations in the heating
radiation field or in the relative abundance of small and large grains
\citep{Laureijs1989, Bernard1992, Ridderstad2010}. Therefore, these
data can give further clues to the nature and the formation process of
the clumps. 

In addition to the SPIRE 250\,$\mu$m data, Fig.~\ref{fig:ids_1} and
the figures in Appendix~\ref{sect:appendix_ids} show the WISE
22\,$\mu$m and the AKARI 90\,$\mu$m, and 140\,$\mu$m surface
brightness maps. For example, the field G4.18+35.79 is apparently
quiescent with no mid-infrared (MIR) or far-infrared (FIR) point
sources (Fig.~\ref{fig:ids_1}). On the other hand, in G130.37+11.26
(Fig.~\ref{fig:ids_1b}) the 22\,$\mu$m data indicate clear star
formation activity in connection with this mainly cold clump. The
enhanced 22\,$\mu$m emission towards the south further suggests an
anisotropy of the radiation field. The lack of correlation or even
clear anticorrelation between the PAH emission dominated MIR and the
FIR/sub-mm emission is a feature repeated in many of the fields.

\begin{figure*}
\centering
\includegraphics[width=16.0cm]{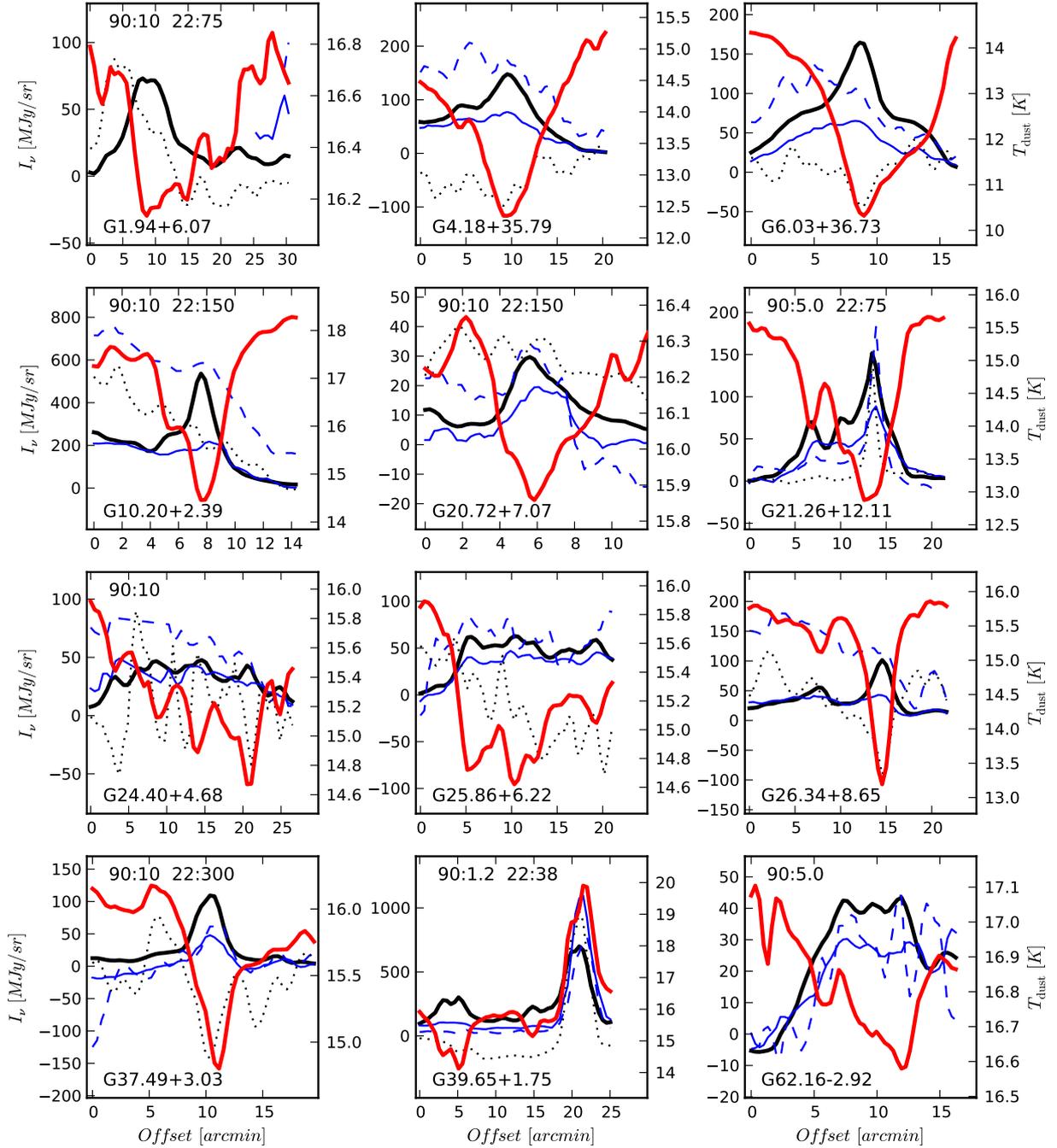}
\caption{
Cross sections of the surface brightness data along the lines indicated in
Figs.~\ref{fig:ids_1}--\ref{fig:ids_1b} and in
the figures of Appendix~\ref{sect:appendix_ids}. The lines show the 250\,$\mu$m SPIRE
data (thick black line), AKARI 140\,$\mu$m and 90\,$\mu$m data (solid and
dashed blue lines), and, when available, the WISE 22\,$\mu$m data (dotted
line). The average surface brightness in the reference region (see
Fig.~\ref{fig:ids_1}b) has been subtracted from the plotted values. The
red thick line is the colour temperature. The data have been convolved to
the resolution of one arc minute. The 90\,$\mu$m data have been scaled by
a factor 20 and the 22\,$\mu$m data by a factor of 600. When a different
scaling has been used, the wavelength and the multiplicative scaling
factor are given in the frame ($\lambda$:factor). The plots for the other
fields are shown in Appendix~\ref{sect:appendix_stripes}.
}
\label{fig:stripes_1}%
\end{figure*}

Figure~\ref{fig:stripes_1} and
Figs.~\ref{fig:stripes_2}--\ref{fig:stripes_6} (online edition) show cross sections of
the surface brightness data along the lines marked in  frame $b$ of
Fig.~\ref{fig:ids_1} and the figures of
Appendix~\ref{sect:appendix_ids}. The figures show the intensities
after the subtraction of the values in the reference regions that were
marked in the Fig.~\ref{fig:ids_1} and the figures of
Appendix~\ref{sect:appendix_ids}.

\subsubsection{Clump properties from dust emission} \label{sect:clumps}

In this section we look in more detail at the properties of some major
clumps, combining {\it Herschel} observations with the infrared data.
The analysis uses the apertures that were marked in the frames $a$ of
Fig.~\ref{fig:ids_1} and figures in Appendix~\ref{sect:appendix_ids} (online edition).

The masses within the apertures were first determined using SPIRE data
only. We derive two mass estimates. The first one is obtained by
directly integrating over the aperture in the column density maps of
Sect.~\ref{sect:colden}. The spatial resolution of the column density
maps is 40$\arcsec$. Because the column densities are calculated using
the absolute surface brightness values, these should correspond to the
total mass along the line of sight. However, the estimates are based
on colour temperatures that, in the case of cold cores, can seriously
overestimate the temperature inside the cores.
We calculate another mass estimate using aperture photometry. The SPIRE surface
brightness maps are convolved to a resolution of 40$\arcsec$. We measure the signal
within the aperture and subtract the background using an annulus that extends to
1.5--2.0 times the radius of the aperture, thus removing most of the diffuse signal.
The aperture sizes are listed in Table~\ref{table:mass}. The uncertainties of the
flux values are derived from the surface brightness fluctuation in the annulus. The
resulting SPIRE fluxes are fitted with a modified blackbody to estimate the clump
masses. Because of the local background subtraction, these estimates should better
reflect the temperature and the mass of the dense structures only. The mass
estimates are listed in Table~\ref{table:mass}.

\begin{figure*}
\centering
\includegraphics[width=14cm]{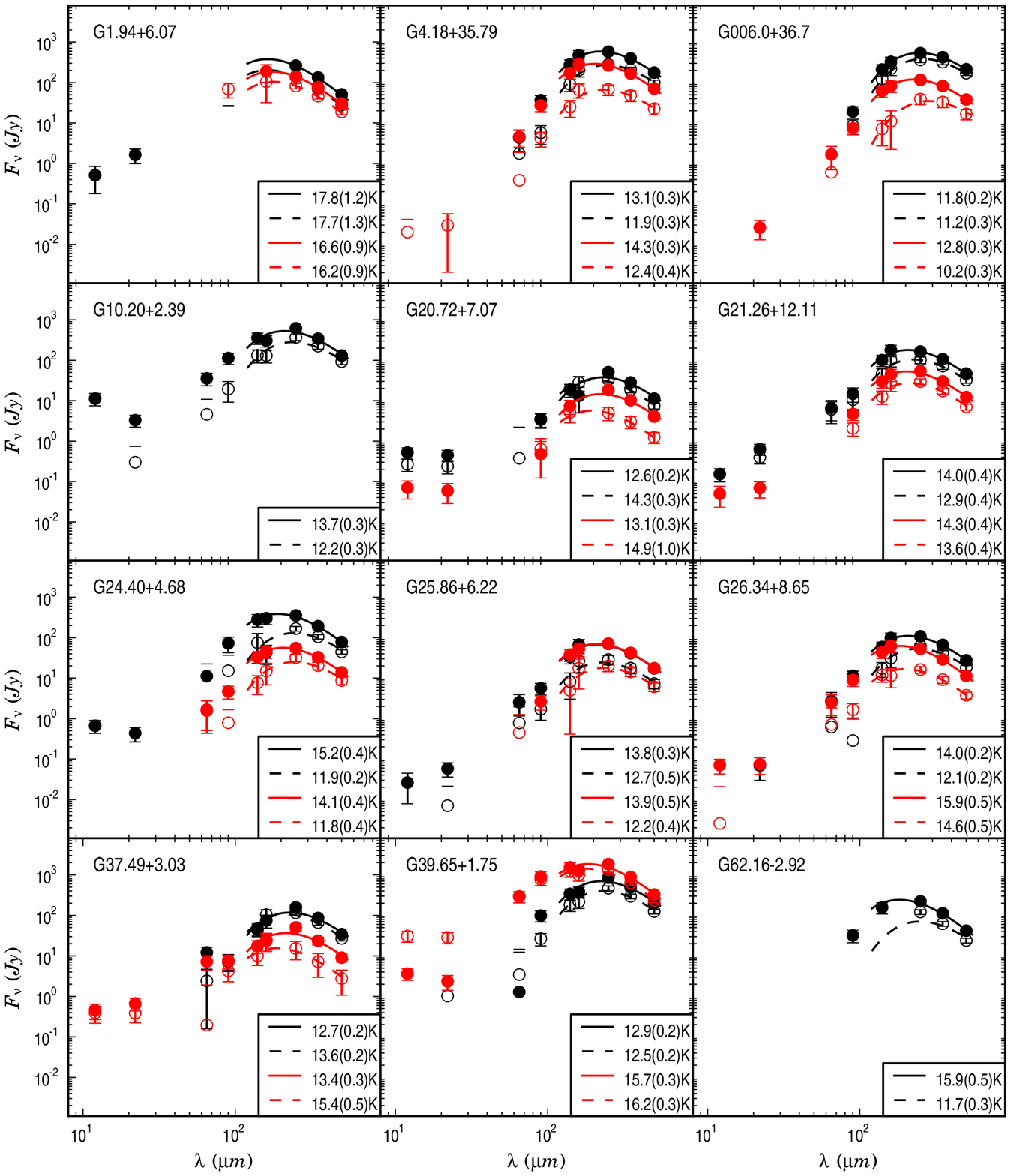}
\caption{
Spectral energy distributions for the apertures indicated in
Fig.~\ref{fig:ids_1} and the Appendix~\ref{sect:appendix_ids}. The plots
include the 22\,$\mu$m WISE data, the AKARI data at 65\,$\mu$m,
90\,$\mu$m, 140\,$\mu$m, and 160\,$\mu$m, and the three SPIRE channels at
250\,$\mu$m, 350\,$\mu$m, and 500\,$\mu$m.
In most fields two aperture positions were chosen and the data for the
second one are shown in red. 
The background subtraction was done using the reference areas marked in
Figs.~\ref{fig:ids_1} and the figures of Appendix ~\ref{sect:appendix_ids}
(solid symbols) or by using a local annulus (open symbols).
The colour temperatures from the modified blackbody fits with $\beta=2$
are listed in the frames.
Plots for the other fields can be found in
Appendix~\ref{sect:appendix_seds}.
}
\label{fig:sed_1}%
\end{figure*}

We have constructed the clump SEDs from 22\,$\mu$m to the SPIRE wavelengths.
Figure~\ref{fig:sed_1} shows the SEDs for clumps in the first twelve fields (in
order of increasing Galactic longitude) and similar figures for the remaining fields
can be found in Appendix~\ref{sect:appendix_seds}. In addition to the three SPIRE
channels (250\,$\mu$m, 350\,$\mu$m, and 500\,$\mu$m) the plots include the AKARI FIS
bands (65\,$\mu$m, 90\,$\mu$m, 140\,$\mu$m, and 160\,$\mu$m) and the WISE data at 12
and 22$\mu$m. The data were convolved to one arcmin resolution. The local background
subtraction was performed as in the case of SPIRE data above, using an annulus
around the aperture. Alternatively, because we do not have absolute surface
brightness measurements in all infrared bands, the background can be estimated using
the reference areas indicated in the frames $b$ of Fig.~\ref{fig:ids_1} and in
Appendix~\ref{sect:appendix_ids}. If the signal in the reference area is small, the
resulting spectra should give a good estimate of the intensity and the spectral
shape of the total emission along the line of sight. The data at wavelengths
$\lambda>100\,\mu$m were fitted with modified blackbodies,
$B_{\nu}(T)\,\nu^{\beta}$, with a fixed value of the emissivity spectral index,
$\beta=2.0$. The resulting colour temperatures are given in the figures. 

\begin{figure}
\centering
\includegraphics[width=8cm]{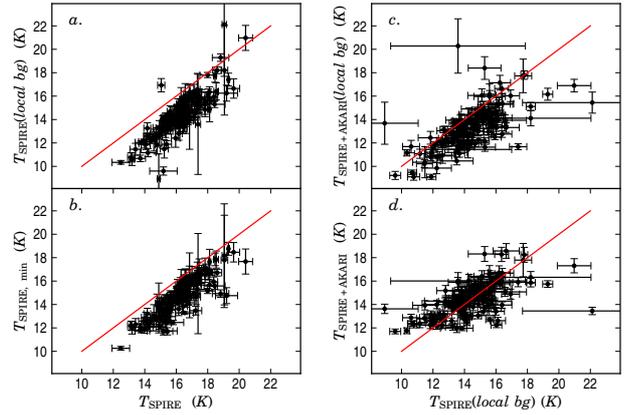}
\caption{
Comparison of the temperature estimates for the selected clumps.  In the
left hand frames, the x-axis is the temperature derived from the total
surface brightness. This is correlated with the minimum values within the
apertures (frame $b$) and the values obtained with a local background
subtraction (frame $a$).
In the right hand frames the SPIRE temperatures (with background
subtraction) are compared with the values obtained using the combination
of SPIRE and AKARI data ($\lambda>100\,\mu$m) and with (frame $c$) or
without (frame $d$) the subtraction of the local background.
}
\label{fig:Tcomparison}
\end{figure}

We now have several colour temperature estimates for each aperture. Two are based on
SPIRE only (as in Fig.~\ref{fig:ids_1} and Table~\ref{table:mass}) and two
additionally include the AKARI 140\,$\mu$m and 160\,$\mu$m data (in
Fig.~\ref{fig:sed_1} and Appendix~\ref{sect:appendix_seds}). Furthermore, in two
cases we use the total surface brightness values (or an approximation obtained with
the help of the reference areas) while in the other two cases the local background
is subtracted using an annulus. The temperature estimates are compared in
Fig.~\ref{fig:Tcomparison}. The subtraction of the local background typically leads
to $\sim$2\,K lower colour temperatures as expected for sources colder than their
environment. The values are lowered by a further 1--2\,K when the FIR data are included in
the modified blackbody fits. One would have expected the temperatures to be lower
when the shorter wavelengths are excluded \citep[see][]{Shetty2009a}. 
The result could suggest a small difference in the calibration of the
two instruments.
The other possibility is that the FIR intensities are partly
saturated. For the typical beam averaged peak column densities of
10$^{22}$\,cm$^{-2}$, the 100\,$\mu$m optical depth is of the order of
0.01, too little to cause noticeable effects. If the mass is
concentrated to a very small area within the beam, the optical depths
will be higher. A 1--2\,K difference in the colour temperature would
require column densities in excess of 10$^{24}$\,cm$^{-2}$. Such
values are possible, for example, in case of distant, pre-high-mass
clumps. However, the optical depths are unlikely to be the main
explanation for the whole sample because that would require that most
of the mass within the beam is always concentrated on such high
optical depth sightlines.

\subsubsection{Search for coreshine} \label{sect:coreshine}

\citet{Steinacker2010} detected enhanced mid-infrared surface
brightness in the Spitzer/IRAC 3.6 and 4.5\,$\mu$m channels towards
the cloud LDN\,183. The high intensity of the 3.6\,$\mu$m band relative
to the 4.5\,$\mu$m band, the absence of emission in the 5.8\,$\mu$m
band, the presence of absorption in the 8 $\mu$m band, and other
considerations, led them to the conclusion that neither PAHs nor warm
grains could explain these observations which were interpreted as a
sign of enhanced light scattering caused by an increase in the size of
the dust grains. This excess, named by the authors as ‘coreshine’,
would thus serve as a tracer of the grain evolution, similar to the
increase of the sub-millimetre dust opacity
\citep[e.g.][]{Stepnik2003, Planck25}.
Thus, this phenomenon could also be used as an indicator of
the evolutionary stage of our clumps. We examined the correlation
between the WISE satellite 3.4, 4.6, and 12\,$\mu$m bands and the
250\,$\mu$m band measurements from SPIRE. If the grain size has indeed
significantly increased, the WISE 3.4 $\mu$m band should include an
additional component of scattered light that makes it brighter than
expected compared to its surroundings and this should be correlated
with the absence of emission, or better the absorption in the WISE
12\,$\mu$m band, weaker or no emission in the WISE 4.6\,$\mu$m band
and these features should appear inside the 250\,$\mu$m emission
region.

The 4.6\,$\mu$m band includes H$_2$ lines that can be strong in
outflows \citep{Cyganowski2011}.  Also the 3.4\,$\mu$m band is
affected by H$_2$ emission but, for typical excitation conditions, the
intensity is much lower than in the 4.6\,$\mu$m band
\citep{Reach2006}. Furthermore, the H$_2$ emission is also seen in the
next MIR bands (5.8\,$\mu$m and 8\,$\mu$m for Spitzer/IRAC and
12\,$\mu$m for WISE) and is therefore easy to distinguish from
coreshine for which absorption is expected at these longer
wavelengths.

\begin{figure*}
\centering
\includegraphics[width=15cm]{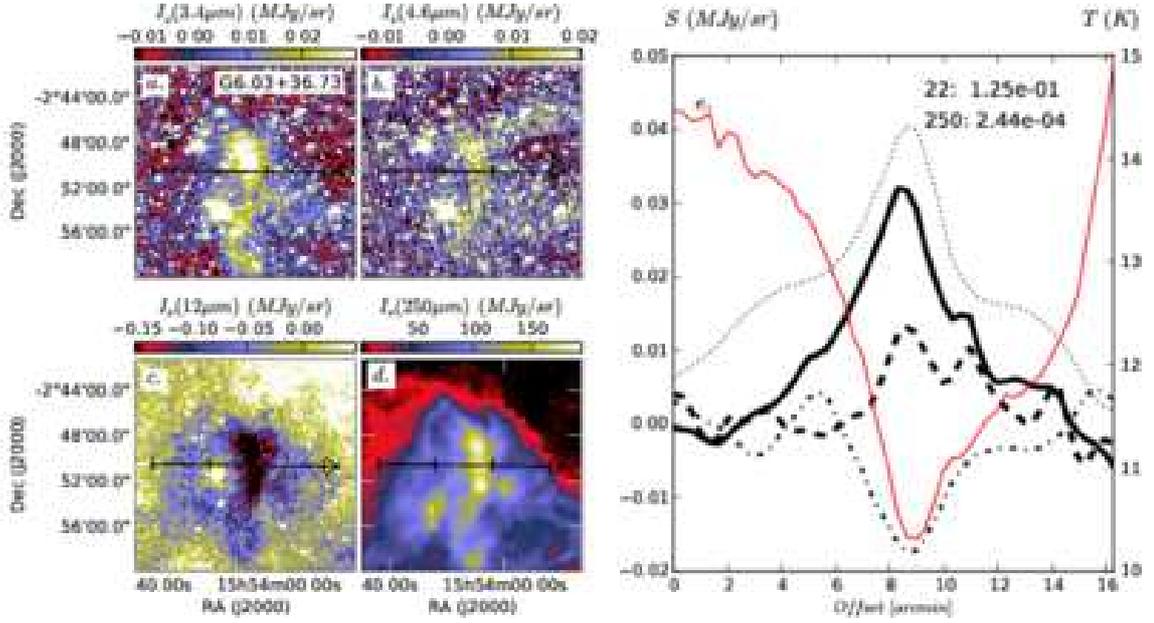}
\caption{
Evidence of coreshine in the field G6.03+36.73 (LDN\,183). Shown are the
WISE 3.4\,$\mu$m, 4.6\,$\mu$m, and 12\,$\mu$m WISE maps (frames $a$--$c$)
and the SPIRE 250\,$\mu$m map (frame $d$). Frame $e$ shows surface
brightness profiles along the arrow marked in the previous frames. The
lines correspond to the 3.4\,$\mu$m (solid line), 4.6\,$\mu$m (dashed
line), 12\,$\mu$m (dash-dotted line), and 250\,$\mu$m data (dotted line).
The numbers in the upper right corner give the possible scaling applied to
the data before plotting ($\lambda$:factor). The red line and the right
hand scale show the colour temperature.
}
\label{fig:coreshine_LDN183}%
\end{figure*}

One caveat concerns the absorption
of the MIR radiation from the sky behind the source. Because the cloud
opacity increases towards the shorter wavelengths, the background
radiation is attenuated more at 3.4\,$\mu$m than at 4.6\,$\mu$m. If
the background level is high, the detection of the coreshine may
require modelling of the absorption and emission processes. This
suggests that the direct effect is best seen at high Galactic
latitudes where the sky brightness is low. This has been remarked by
\citet{Pagani2010} who could find no cases of coreshine emission in
the galactic plane itself. An upper limit of 0.37\,MJy\,sr$^{-1}$ in
the background brightness level (which is linked to the density of the
field stars) to allow for coreshine detection has been derived and
will be discussed elsewhere (Pagani et al. in preparation). Another
limitation comes from the low resolution of the WISE satellite
(6.1\arcsec\ for all three bands, 3.4, 4.6 and 12\,$\mu$m), which,
combined with a high density of stars close to the galactic plane, tends
to fill up the space with starlight, hiding possible coreshine
effects. Figure~\ref{fig:coreshine_LDN183} shows data for the field
G6.03+36.73 (LDN183), one of the strongest coreshine detections,
already discussed by Steinacker et al. (2010). The central filament,
as traced by the 250 $\mu$m emission, is clearly visible in the
3.4\,$\mu$m map. There is clear detection in emission at 4.6\,$\mu$m
and in absorption at 12\,$\mu$m. The coreshine intensity is comparable
to the one reported by \citet{Steinacker2010}. The 3.4\,$\mu$m signal
increase is larger than for the 4.6\,$\mu$m one and the cross-sections
of the surface brightness show that the increase of the
3.4\,$\mu$m/4.6\,$\mu$m ratio is correlated with both the 250 $\mu$m
emission and the decrease of the dust temperature.

\begin{figure*}
\centering
\includegraphics[width=15cm]{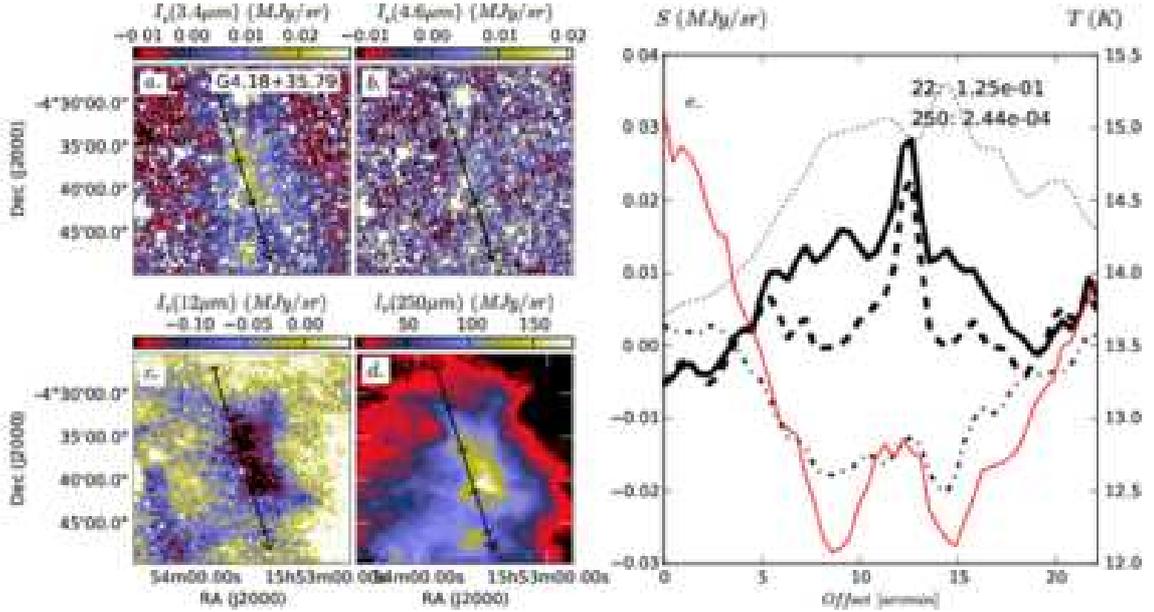}
\caption{
Evidence of coreshine in the field G4.18+35.79 (LDN\,134). See
Fig.~\ref{fig:coreshine_LDN183} for a description of the figure.
}
\label{fig:coreshine_LDN134}%
\end{figure*}

We have similarly examined the 3.4, 4.6 and 12\,$\mu$m images of all
the 56 fields for which public WISE data are available. We compare
them with the 250\,$\mu$m SPIRE surface brightness maps that act as a
tracer of the large grains.
Figure~\ref{fig:coreshine_LDN134} shows another example of detected
coreshine.

\section{Discussion}  \label{sect:discussion}

\subsection{The nature of the observed fields}

The observations show that the clouds associated with cold clumps have varied
morphologies. The cloud sizes range from nearby globules $\sim$0.2\,pc  to massive
filaments more than 20\,pc in length. With very few exceptions, the clumps exhibit
a significant amount of sub-structure down to the resolution limit (0.01\,pc for
the nearest regions, over 0.2\,pc for the most distant ones). 

Of the 71 fields examined, 36 fields or 51\% were categorized as having
a clear filamentary structure.  The definition of a `filament' here
applies to the major structures but is taken in a widest possible sense,
meaning anything from slightly elongated structures (e.g., G1.94+6.07 or
G10.20+2.39) to the very long and narrow `real' filaments like those
observed in the fields G82.65-2.00 and G276.78+1.75. Several filaments
appear to have formed, or at least significantly deformed, by some external
force. This is visible as an asymmetry or a sharp drop in the spatial
distribution of emission and column density (e.g., G25.86+6.22,
G94.15+6.50, G130.37+11.26, or G315.88-21.44, see
Figs.~\ref{fig:ids_06}, \ref{fig:ids_17}, \ref{fig:ids_1b}, and
\ref{fig:ids_64} in online edition). Thus, the origin of these filamentary structures
could be different, i.e., turbulence vs. direct compression by radiation
or pressure waves. 
This could be reflected in their properties which are further discussed in
Sect.~\ref{sect:discussion_filaments}.
However, further study of the interaction with, e.g., nearby HII regions is
deferred to a future paper.

Altogether eight fields showed features reminiscent of compression by
an external force (marked with `B' in column 8 of
Table~\ref{table:main}) that has lead to a sharp boundary visible in
the column density maps. Such an interaction can also manifest itself
at larger scales, as a cometary of a filamentary shape of the entire cloud.
This sample includes some well known cometary shaped clouds (e.g.,
LDN\,1780). At low resolution, the cloud LDN1340 (field G130.37+11.26)
also looks like a classic cometary cloud but at higher resolution is
found to consist of an intricate network of filaments that,
particularly in the tail of the cloud, breaks down to many smaller
clumps.

The clumps with a simple structure are mostly nearby objects where the
spatial resolution allows us to resolve the core scales. However,
there also are some more distant objects in this category. The target
G39.65+1.75 is at a distance of 1.8\,kpc and it is probable that it
would not be such a featureless `blob' if it could be better resolved. On
the other hand, the main clump in G26.34+8.65 is resolved but shows
remarkably little structure for such a distant and therefore large
object. However, one must keep in mind that the distance estimates of
some of the objects are still quite uncertain and the distances of
objects like G26.34+8.65 might be overestimated.

The colour temperature maps (Figs.~\ref{fig:ids_1}--\ref{fig:ids_1b}
and Appendix~\ref{sect:appendix_ids}) show that the dense structures
are, in colour temperature, typically 2--3 degrees colder than their
diffuse environment. The appearance of the temperature maps, i.e.
their strongly spatially correlated values, indicates that the maps
are dominated by real temperature variations rather than noise.
The analysis of the uncertainties suggested that in the temperature
maps the relative values are quite reliable
(Sect.~\ref{sect:temperature}) although calibration errors could cause
a shift of about $\pm$1\,K (see Sect.~\ref{sect:clumps}).

\begin{figure}
\centering
\includegraphics[width=8cm]{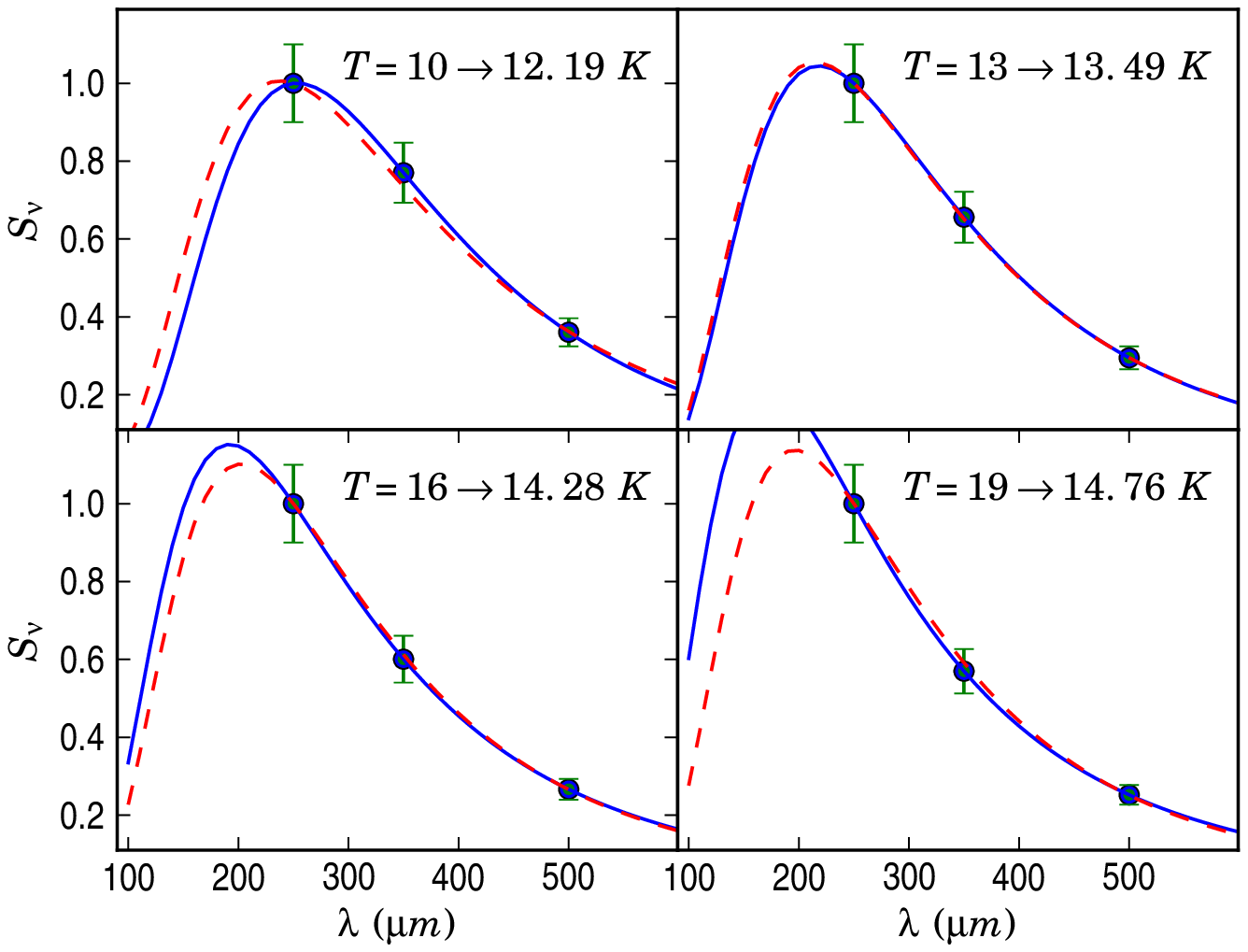}
\caption{
Spectral energy distributions (solid curves) for 10, 13, 16, and 19\,K
colour temperatures assuming the $\beta(T)$ relation given in
\cite{PlanckI}. The corresponding values at the SPIRE wavelengths are
plotted with 10\% error bars. The dashed lines show the spectra fitted to
these data points assuming $\beta=2.0$. The corresponding temperature
values are quoted after the arrows.
}
\label{fig:beta_test}%
\end{figure}

The colour temperature calculations were based on the use of a fixed
value of the dust emissivity spectral index, $\beta$. There is
evidence that, at least statistically, there exists an inverse
relation between the observed spectral index and the dust temperature
\citep{Dupac2003, Desert2008, PlanckI, Veneziani2010, Paradis2010}. 
It is still debated to what extent this is caused by the
intrinsic dust properties, by the line-of-sight temperature variations
\citep{Shetty2009a, Malinen2011, JuvelaYsard2012a, YsardJuvela2012}, and, in particular, by the
noise that is known to produce some anticorrelation between the colour
temperature and the spectral index \citep{Shetty2009b,
JuvelaYsard2012b}.
However, the anticorrelation appears to be an established
observational fact and stronger than what is produced by the
noise only \citep{PlanckI}. 
If the average spectral index is larger in cold regions, the
temperature contrasts (i.e., the difference in derived colour
temperature) would be higher than indicated by our maps.
Figure~\ref{fig:beta_test} shows some spectra consistent with the
relation $\beta(T)=(\delta + \omega T)^{-1}$ with parameters
$\delta=0.02$ and $\omega=0.035$ as estimated from the analysis of
Planck observations of dense cores \citep{PlanckI}. The relative
intensities in the SPIRE channels and the SED fits obtained with a
fixed value of $\beta=2.0$ are shown. For example, when a value of
$\beta=2.0$ is assumed, the estimated colour temperature is 12.2\,K
while the actual values of the spectral index and temperature are
$\beta$=2.7 and $T_{\rm dust}$=10\,K. The figure also illustrates
that, without further far-infrared data, simultaneous determination of
colour temperature and spectral index is not possible.

We conclude that the colour temperature maps do reliably identify the coldest clumps
although the absolute temperature values may be uncertain up to a couple of degrees.
The real dust temperature inside the clumps cannot be directly measured and could be
estimated only by modelling. From earlier theoretical studies, it is clear that in
many cases the central temperature of the dense cores goes below 10\,K \citep[e.g.,
][]{Evans2001, Harju2008}. In particular, in LDN~183, our field G6.03+36.73,
\cite{Pagani2004} reported dust temperatures down to $\sim$7\,K.
In the dense and cold
environment the dust grains are expected to undergo coagulation that
can lead to increased opacity at sub-millimetre wavelengths
\citep{Ossenkopf1993, Stepnik2003, YsardJuvela2012, Kohler2012}. 
At the same time,
the abundance of small grains should decrease, leading to lower
mid-infrared emission. Examination of the data for the apertures shown
in the figures of Appendix~\ref{sect:appendix_ids} revealed no
significant anticorrelation between the surface brightness ratio
22\,$\mu$m/90\,$\mu$m and the ratio
90\,$\mu$m/250\,$\mu$m or the colour temperature measured in the
wavelength range 250--500\,$\mu$m. There could be several reasons for
this. Firstly, most clumps could be too young so that the coagulation
process, which may take millions of years, has not had time to alter
the abundance of the small grains \citep{Ossenkopf1994,
Chakrabarti2005}. Another natural reason is that the 22\,$\mu$m
emission comes mainly from a region with $A_{\rm V}< 2^{\rm m}$, as
indicated by MIR limb brightening in radiative transfer models
\citep{Bernard1992, Fischera2008}, and thus is not sensitive to
processes deep inside the cores.
The IR point sources associated with the cold clumps are another
complication. Many regions are already forming stars so that the MIR
data, the 22\,$\mu$m band included, are dominated by the warm sources
rather than the extended dust emission. This contaminates the
correlations between MIR emission and the mean colour temperature of
the large grains.

In a few individual clumps, an anticorrelation does exist between the
column density, as traced by the sub-millimetre emission, and the MIR
emission. These include the fields G6.0+36.7, G37.49+3.03, and
G26.34+8.65 as shown in Fig.~\ref{fig:stripes_1} and some further
examples can be seen in Appendix~\ref{sect:appendix_stripes} (online edition).

\subsection{Connection with star formation} \label{sect:starformation}

\begin{figure*}
\centering
\includegraphics[width=17cm]{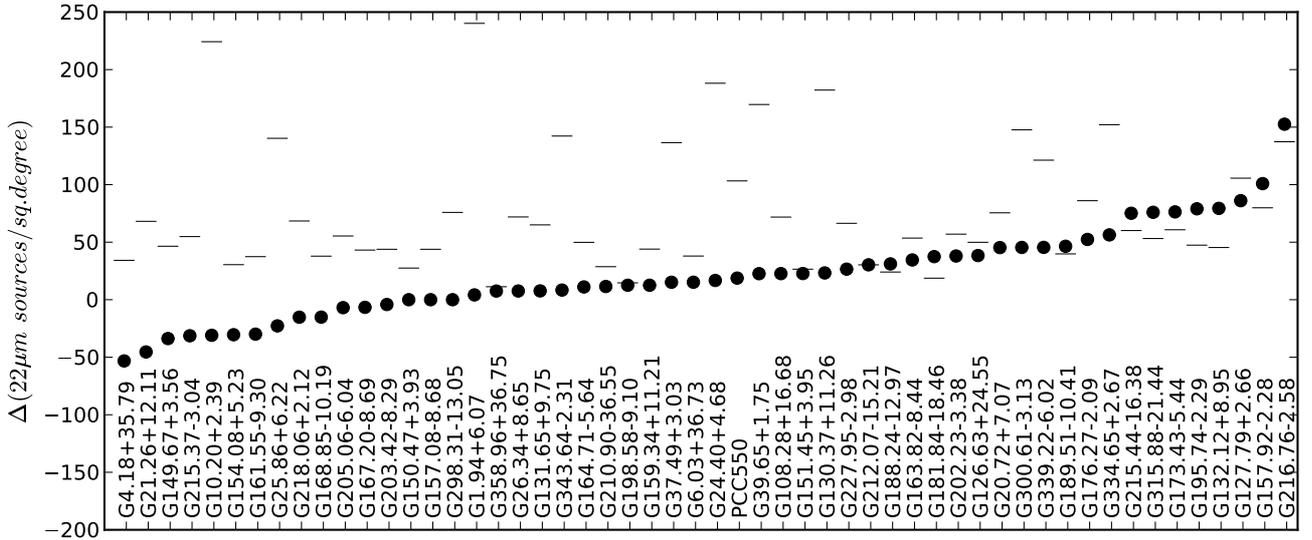}
\caption{
The fields sorted in increasing number density of the 22\,$\mu$m
sources in the high column density areas relative to the average source
density of the field. The circles are the excess of sources per square
degree. The horizontal dashes show the total source density, including
both the high and the low column density areas.
}
\label{fig:SF_correlation}%
\end{figure*}

Star formation activity is visible in the MIR maps as sources
spatially correlated to the cold dust emission. The best example is
the field G216.76-2.58 where ten strong WISE MIR sources are detected
along the high column density structures. This does not prevent cold
dust being detected at longer wavelengths in exactly the same areas,
with colour temperatures down to $T_{\rm dust}\sim 12$\,K.
As a preliminary quantitative estimate of the star formation activity
we counted the surface density of MIR sources. For each field for
which WISE data were available, we made a list of the 22\,$\mu$m point
sources. The detection was done with the SExtractor program, excluding
sources that are significantly more extended than the beam (FWHM
larger than 10 pixels = 60$\arcsec$). Each field was divided into a low
and a high column density part using the median column density as the
dividing value. The source densities were calculated for the two
parts. Figure~\ref{fig:SF_correlation} shows the source density in the
high column density areas (i.e., sources likely to be associated
with the clumps) relative to the average source density in the field.

The interpretation of such correlations is not straightforward. The
youngest objects should reside within the areas of high column density
and thus lead to a positive correlation (fields on the right hand side
of Fig.~\ref{fig:SF_correlation}). A negative correlation could appear
in more evolved regions where the sources have cleared their
surroundings. However, with the typical clump column densities of
10$^{22}$\,cm$^{-2}$ ($A_{\rm V}\sim 10$), the 22$\mu$m optical depth
approaches unity and this is enough to affect the number density of
the detected background sources. Therefore, the fields in the left
hand part of Fig.~\ref{fig:SF_correlation} could represent either
cases without associated star formation or cases with a larger number
of background sources. 
The number of MIR sources and the correlations may be further
affected by the source distances because of the sensitivity limit of
the WISE survey.
However, the overall source densities of the fields and
the fraction of sources in the high column density areas do not appear
to be correlated (see Fig.~\ref{fig:SF_correlation}).

About 50 percent of the clumps are located on FIR loops
\cite{Konyves2007} that are boundary regions (shells) around the voids
in the galactic ISM. A preliminary analysis of a sample of sources and
their AKARI FIR photometry showed that the clumps located on the
shells have $\sim$20\% higher probability of being associated with YSO
candidates 
\citep{Marton2012}.
In particular, a detailed analysis of the 62 IR point sources in the
``snake'' cloud (California nebula, G163.82-8.44) region resulted in a
number of YSO candidates. The sources were analysed using the AKARI IRC
and FIS \citep{Ishihara2010, Murakami2007, Yamamura2010}, the 2MASS and
IRAS point source catalogues, as well as optical photometry data
available in public archives. On the basis of the physical parameters
derived for the sources, there are 11 low-mass and 2 intermediate mass
young stellar objects in evolutionary phases ranging from Class 0/I to
Class III. In that particular field, the estimated ages suggest that the
star formation started several million years ago and is still on-going.
Further details will be given in a separate paper (Zahorecz et al. 2011
in preparation).

\subsection{Filamentary structures} \label{sect:discussion_filaments}

\subsubsection{The uncertainties of filament analysis}

The characterisation of the filaments depends on the way the analysed structures are
selected and on the analysis methods themselves.  We selected 
at most one elongated structure  from each field. One should consult the figures in
Appendix~\ref{sect:appendix_filaments} to check to what extent these correspond to
ones own conception of filaments. There are automatic routines for the detection of
filaments, for example the DisPerSE routine \citep{Sousbie2011} used in
\cite{Arzoumanian2011} and \cite{Hill2011}. We chose to do the selection manually,
limiting the analysis to the main structures. Once these have been selected, we
followed the ridge of the filament, connecting the column density peaks sampled at
20$\arcsec$ intervals. Unless the filament has a very well defined, smooth ridge,
 this decision affects the derived filament widths. The structures are often
fragmented and the column density peaks do not necessarily follow a single line. If
one draws a wiggly filament through every individual clump, one is measuring the
sizes of those smaller structures. One could equally describe the system as a more
straight filament, allowing the spatial scatter of the substructures to be counted
as part of its width. Another critical parameter is the spatial extent analysed. For
the FWHM values the effect is obvious and the values increase with the size of the
analysed area, i.e., with the maximum distance from the centre of the filament. This
is caused by the hierarchical structure of the ISM and the need to apply background
subtraction at some scale to remove the contribution of extended emission that is
not directly connected with the filament. The fit of the Plummer profile may be less
affected, but only as far as the actual column density profile can be precisely
fitted with that particular functional form.
Because of these ambiguities, the width of the filament and its other parameters
are not well defined and different studies can be compared only when the methods and
the probed linear scales are identical. In particular, to be able to compare sources
at different distances, we were forced to select $\pm$0.4\,pc as the extent of the
structures examined.

There are several reasons why our dust observations may give an incomplete picture
of the structure of the clouds. Firstly, the column density values may be biased in
the presence of temperature variations that exist along the line of sight and within
the beam. Because warm dust emits more strongly, the colour temperature
overestimates the true, mass weighted dust temperature. Consequently, the column
densities will be underestimated and the effect is most notable towards the centre
of the filaments where the temperature variations along the line of sight are the
largest. In our analysis of Sect.~\ref{sect:filaments} we use column density maps
that are based on lower, 40$\arcsec$ resolution colour temperature information. To
quantify the associated biases, we examined radiative transfer models of cylinders
with Plummer like density profile \citep{YsardJuvela2012}. The dust
temperature distributions were solved assuming that the cylinders are heated
externally by an interstellar radiation field with intensity applicable to the solar
neighbourhood \citep{Mathis1983}. The resulting synthetic surface brightness maps
were analysed as the SPIRE observations, deriving colour temperature maps at the
40$\arcsec$ resolution and the column density maps at the 20$\arcsec$ resolution.
The data were then fitted with Plummer functions and the obtained parameters were
compared to the actual values of the column density distribution.
For a filament $\rho_{\rm C}=10^4$\,cm$^{-3}$, $R_{\rm flat}$=0.58\,pc, and $p=2.0$,
the correct parameters were recovered with an accuracy better than 10\%, even when the
source was moved to a 1\,kpc distance and a 3\% observational noise was added to the
surface brightness values. The relative errors increase with the density of the
filament, also because the density peak is less well resolved. With $\rho_{\rm
C}=10^6$\,cm$^{-3}$, $R_{\rm flat}$=0.0055\,pc, and $p=2.0$, the parameter
$\rho_{\rm C}$ was underestimated by $\sim$30\% and $R_{\rm flat}$ similarly
overestimated by $\sim$30\% for a source at 100\,pc. For a more distant
target with (1\,kpc), the typical error was a factor of three larger.
The
estimates of the parameter $p$ still remain correct to within 30\% in all the tests.
The results suggest that the central density will be underestimated and the filament
radius, $R_{\rm flat}$, overestimated but, for the typical filaments in our sample,
the errors are below 50\%. In the tests the density of the filaments perfectly
followed the Plummer profile. Therefore, it does not indicate if the results would
be sensitive to small deviations from the Plummer shape such as, e.g., in the
presence of asymmetries or substructure. However, it is clear that the $\rho_{\rm C}$
and $R_{\rm flat}$ parameters are intrinsically anticorrelated.

Further studies are required to find out how the combination of
low resolution colour temperature and higher resolution surface
brightness data compares with the normal procedure of convolving all
data to the lowest common resolution.

\subsubsection{Filament structure}

The width of the filaments (or more generally, of the selected
structures shown in Appendix~\ref{sect:appendix_filaments}) were found
to cover a large range from $\sim$0.1pc to close to 1 parsec. 

The FWHM widths are often larger than 0.1\,pc and also vary along the
length of the filaments. The width is often anticorrelated with the
column density, the smallest FWHM values being associated with the
highest column densities, i.e., dense clumps and cores. Part of this
anticorrelation is caused by the non-continuous nature of some
filaments where no real column density excess is detected between the
clumps.  Depending on the structure of the surrounding ISM, the widths
obtained in the gaps between the clumps can be very large. However,
there is also real anticorrelation that can be seen in fields like
G300.86-9.00 (PCC550), G167.20-8.69, and G227.95-2.98.  

The distances between denser cores can be as low as $\sim$0.1\,pc
(e.g. G110.89-2.78) but are typically 0.5\,pc or higher and thus
significantly above the Jeans length. For example, G167.20-8.69 shows
a very regular pattern with column density peaks at 0.6\,pc intervals.
The results are consistent with \citet{Henning2010} who measured a core separation of
0.9\,pc  in the infrared dark cloud filament G011.11-0.12.

\begin{figure*}
\centering
\includegraphics[width=15.5cm]{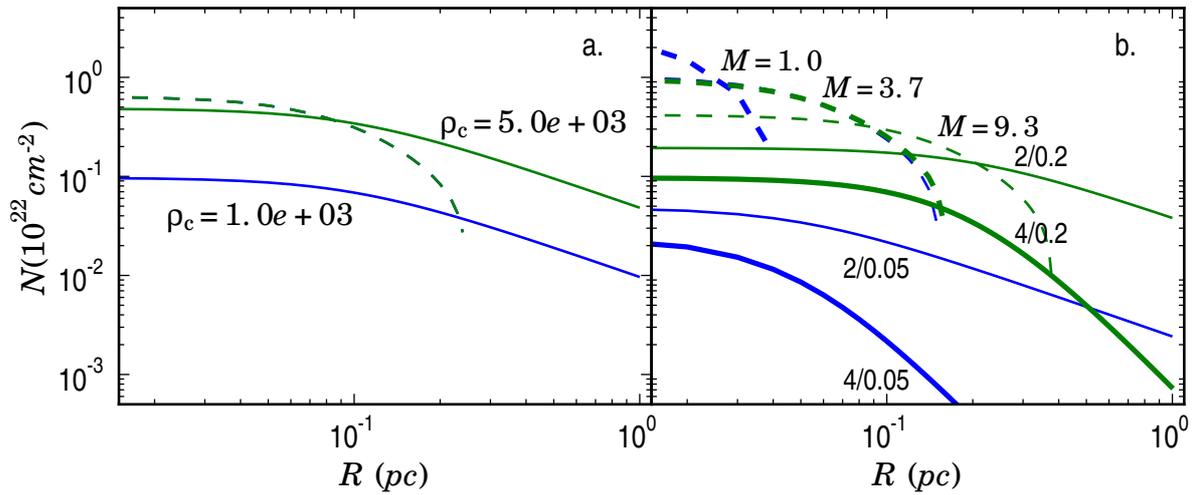}
\caption{
Radial column density profiles of selected Plummer profile filaments
(solid lines) and Bonnor-Ebert spheres (dashed lines). {\em a)} Two
filaments with $R_{\rm flat}=0.1$\,pc and $p=$2 and a central density of
$\rho_{c}=10^3$\,cm$^{-3}$ or $\rho_{c}=5\cdot 10^3$\,cm$^{-3}$, and a
Bonner-Ebert sphere with a mass of 6.0$M_{\sun}$ that in both cases
corresponds to the filament mass within one Jeans length.
{\em b)} Filaments with a central density $\rho_{c}=10^3$\,cm$^{-3}$ with
values $p$/$R_{\rm flat}$ as indicated in the figure. The dashed lines are
column density profiles for Bonnor-Ebert spheres with mass equal to the
mass of the filaments within a distance of one Jeans length.
}
\label{fig:filament_core_widths}%
\end{figure*}

Several fields are actively forming stars and many of the observed
dense cores are likely to be gravitationally bound as discussed in
\cite{PlanckII}. This strongly suggests that many of the mapped cores
are undergoing gravitational collapse. However, anticorrelation
between column density and filament width is expected even for 
isothermal gas cylinders in hydrostatic equilibrium. Similarly, when
hydrostatic spherical clumps reside inside hydrostatic filaments, the
former are likely to have smaller FWHM values.
Figure~\ref{fig:filament_core_widths} shows Plummer profiles and the
column density profiles for Bonner-Ebert spheres with masses equal to
the filament mass within one Jeans length (assuming $T$=10\,K). The
cores are seen to have higher column densities and smaller FWHM
values. Note that in this calculation, if the central density of the filament is very high,
the peak column density of the corresponding Bonnor-Ebert sphere could
 remain lower because of the very short Jeans
length. The figure also illustrates that if the filament profiles extend
far beyond the 0.4\,pc distance used in our analysis, the FWHM
estimates (as opposed to the values obtained from the profile fits)
will be biased toward lower values.

A few sources (G89.65-7.02, G159.34+11.21, G176.27-2.09, G276.78+1.75)
could not be fitted properly using the limited range of $\pm$0.4\,pc. In
most cases this results from the fact that one is looking at some
elongated structures whose basic nature is different from that of actual
filaments. However, the fields G176.27-2.09 and G276.78+1.75 clearly do have
the appearance of regular filaments while their widths are closer to
1\,pc instead of the more typical $\sim$0.2--0.3\,pc. 
In both cases the estimated distances are large, $\sim$2\,kpc. This
again raises the question whether these could in fact be more nearby
clouds.

The widths of our filaments are on the average larger than those
reported by \cite{Arzoumanian2011} for the filaments in IC5146. In
some cases our structures might be further divided into smaller
filaments, depending on the way the filament detection is performed
\citep[cf.][Fig. 1]{Arzoumanian2011}. Thus, clouds can exhibit a
hierarchical structure also regarding the filaments.

\subsection{Coreshine}

Out of the 56 fields examined for the presence of coreshine, 12 are
too close to the galactic plane and show background emission above the
threshold. They are not considered any further. At 12\,$\mu$m, 9
fields showed (weak) emission (most probably linked to PAH emission),
13 fields were not detected and 22 showed absorption. Emission at
3.4\,$\mu$m was found either in correlation with the emission at
12\,$\mu$m (PAHs) and in a few cases in correlation with absorption at
12\,$\mu$m (coreshine). Apart from the well-known L183 case
(G6.03+36.73), coreshine was clearly identified in three other fields
only, G4.18+35.79 (LDN134), G210.90-36.55 (LDN1642, MBM20-21) and
G300.86-9.00 (PCC550, in the Musca Complex). LDN\,134 was erroneously
reported as showing no coreshine in \cite{Pagani2010}. LDN\,1780
(G358.96+36.75), the third cloud in the same complex as LDN134 and
LDN183, is bright at 12\,$\mu$m, unlike the two others and is
therefore dominated by PAH emission. There are six additional fields
with clumps showing possible signs of coreshine (G130.37+11.26,
G167.20-8.69, G173.43-5.44, G181.84-18.46, G215.44-16.38,
G298.31-13.05). However, the MIR intensity ratio is seen to have large
variations that are not correlated with the dust column density (or the
dust temperature). Therefore, these detections can be said to be only
tentative. The sample also includes targets like G39.65+1.75 where the
relative intensity of the 3.4 $\mu$m and 4.6 $\mu$m appears to be
determined mainly by the extinction of background radiation. Most of
the fields with tentative or positive coreshine detection are at high
latitudes. Although not an absolute requirement, a low background
surface brightness makes it easier to detect the faint MIR excess
caused by the scattering. In many other medium to high latitude
sources (e.g., G108.28+16.68, G212.07-15.21, G315.88-21.44) the
results were negative or inconclusive. Finally, only 7 to 18\% of the
fields show confirmed or possible coreshine. This is markedly
lower than the results reported by \citet{Pagani2010} with a
$\sim$50\% detection rate. 

It is not yet understood why some cores show the coreshine effect
while others do not.
Time is needed to let the grains grow but time is
clearly not the only factor. A high peak column density also is not a
necessary nor a sufficient criterion as coreshine is clearly detected
towards sources with N(H$_{2,peak}$) $\leq
1\,\times\,10^{22}$\,cm$^{-2}$ and still not towards sources at
similarly high latitude but with peak column densities 2--3 times as
high. The low resolution of the WISE experiment is possibly a strong
limitation, combined with the small extent of many of the clumps.
The coreshine studies would benefit not only from higher spatial
resolution but also from data with a higher signal-to-noise ratio. The
Spitzer program {\em Hunting Coreshines with Spitzer} (PI R. Paladini)
is currently carrying out observations that, compared to the WISE
data, will be more sensitive by a factor of ten. The targets of that
survey, 90 in number, have been selected from the Planck Early Cold
Cores Catalogue (ECC) and correspond to $\sim$10\% of the ECC. 
Detailed modelling is needed to separate the coreshine from
the effects of the attenuation of the background radiation and to
quantify the degree of the grain growth.

\section{Conclusions}  \label{sect:conclusions}

We have examined a sample of 71 fields that were mapped with the {\it
Herschel} SPIRE instrument as part of the key programme {\em Galactic
Cold Cores}. The examination of the sub-millimetre {\it Herschel}
observations and the available AKARI and WISE infrared data leads to
the following conclusions:
\begin{itemize}
\item The data confirm the presence of cold dust with colour temperatures
of the total intensity typically going down to $\sim$14\,K or below.
With the subtraction of the local background emission, the estimated
temperature of the larger clumps can be at least 1--2\,K lower, the effect being larger if
one assumes that the dust emissivity spectral index in these regions
rises above $\beta=2.0$. 
\item About 50\% of the fields have a mainly filamentary structure,
the filaments often being fragmented into a number of clumps.  In
several cases the morphology belies direct dynamic interaction, e.g.,
in the form of sharp interface layers. Several clearly cometary clouds
are also seen, again combined with a significant amount of smaller
scale structures. 
\item The fields include a few quiescent clouds but most clouds
show some signs of ongoing star formation. Young stellar objects, as
detected in mid-infrared data, are found to co-exist with clumps having
cold sub-millimetre spectra.
\item The mid-infrared data were searched for signs of coreshine, enhanced
mid-infrared scattering potentially caused by the increased grain
sizes. Clear signal was observed in four fields with a further six
tentative detections.
\item The main filamentary structures found in the fields were analysed. The typical
filament width was observed to be $\sim$0.2--0.3\,pc. The widths were
anticorrelated with the column density, a behaviour expected both for
isothermal, hydrostatic filaments of different densities and for
gravitationally bound spherical cores embedded in filaments.
\item Statistically, no anticorrelation was found between the mid-infrared
emission and the large grain temperature, as expected in the dust
coagulation scenario. This can be explained by the fact that these
wavelengths probe entirely separate parts of the clumps. On the other
hand, several clumps were opaque enough to be seen as mid-infrared
dark clouds, with a sharp decrease in the 22\,$\mu$m surface
brightness.
\end{itemize}

\begin{acknowledgements}
MJ, JM and NY acknowledge the support of the Academy of Finland Grants
No. 127015 and 250741. NY acknowledges the support of a CNES
post-doctoral research grant.
This publication makes use of data products from the Two Micron All Sky
Survey, which is a joint project of the University of Massachusetts and
the Infrared Processing and Analysis Center/California Institute of
Technology, funded by the National Aeronautics and Space Administration
and the National Science Foundation.
This research is based in part on observations with AKARI, a JAXA project
with the participation of ESA.
\end{acknowledgements}

\bibliography{biblio_v1.3}

\clearpage
\onecolumn

\renewcommand{\footnoterule}{}

\begin{longtable}{lllllllll}
\caption{
The observed fields. The columns give the field name, the centre
position in Galactic and equatorial coordinates, the size of the
mapped area, the distance, flags for the source morphology, and a list
of possible other identifications.
} \label{table:main} \\
\hline\hline            
Field                                        & 
\multicolumn{2}{c}{Galactic coordinates}     &
\multicolumn{2}{c}{Equatorial  coordinates}  &
Size                                         &
Distance                                     &
Cat.                                         &
Other identifications                       \\
                       &
(deg) & deg)           &
(J2000) & (J2000)      &
($\arcmin$)            &
(kpc)                  &
                       \\
\hline                       
\endfirsthead
\caption{continued.} \\
\hline \hline
Field                                        & 
\multicolumn{2}{c}{Galactic coordinates}     &
\multicolumn{2}{c}{Equatorial  coordinates}  &
Size                                         &
Distance$^1$                                 &
Cat.                                         &
Other identifications                    \\
                       &
(deg) & deg)           &
(J2000) & (J2000)      &
($\arcmin$)            &
(kpc)                  &
                       \\
\hline
\endhead
\hline
\endfoot
     G1.94+6.07-1 &    2.06 &   5.93 & 17 28 07.6 & -24 01 27.7 &   49  & 0.15 (5)
\footnote{
The reference for the distance estimate is given in parentheses: 
 1 -- association with \citet{Simon2006a} IRDC, 
 2 -- reddening of SDSS stars \cite[see][]{McGehee2012}),
 3 -- 3D extinction mapping using 2MASS catalogue stars 
      \cite[see][]{Marshall2009},
 4 -- kinematic distance, 
 5 -- association with a known molecular cloud complex, 
 6 -- direct identification with an object with a distance estimate available in the literature.}     
     &  F
     \footnote{Characterisation of the field morphology:
     F -- filamentary,
     B -- boundary,
     C -- cometary,
     I -- isolated,
     X -- complex,
     }
       &  B77, LDN69 \\       
      G4.18+35.79 &    4.18 &  35.79 &  15 53 31.0 &   -04 37 18 &   37  & 0.14 (6) &     S &  LDN134, MBM36, MLB40 \\
      G6.03+36.73 &    6.03 &  36.73 &  15 54 14.0 &   -02 52 10 &   37  & 0.11 (6) &     I &  LDN183, MBM37 \\
      G10.20+2.39 &   10.20 &   2.39 &  17 59 20.2 &   -18 56 09 &   37  & -        &     I &   \\
      G20.72+7.07 &   20.72 &   7.07 &  18 03 38.7 &   -07 30 23 &   37  & -        &    FI &   \\
     G21.26+12.11 &   21.26 &  12.11 &  17 46 55.8 &   -04 36 47 &   37  & 0.73 (3) &     I &  LDN425, LDN428, LM240 \\
      G24.40+4.68 &   24.31 &   4.56 &  18 19 21.3 &   -05 32 50 &   50  & -        &     X &  LDN475, LDN477, LDN470 \\
      G25.86+6.22 &   25.86 &   6.22 &  18 16 20.4 &   -03 24 52 &   37  & -        &   FBX &  LDN500 \\
      G26.34+8.65 &   26.34 &   8.65 &  18 08 37.6 &   -01 51 29 &   37  & 0.96 (3) &     I &  LDN502, CB112, P61 \\
      G37.49+3.03 &   37.49 &   3.03 &  18 48 56.7 &   +05 26 06 &   37  & -        &     S &  BDN31.48+3.02 \\
      G39.65+1.75 &   39.61 &   1.86 &  18 57 00.5 &   +06 47 17 &   43  & 1.82 (3) &     X &   \\
      G62.16-2.92 &   62.17 &  -2.93 &  19 59 47.0 &   +24 15 26 &   37  & 1.11 (3) &     I &   \\
      G69.57-1.74 &   69.71 &  -1.70 &  20 13 25.5 &   +31 16 20 &   43  & 1.58 (3) &     S &   \\
      G70.10-1.69 &   70.22 &  -1.50 &  20 13 58.1 &   +31 48 36 &   61  & -        &     X &   \\
      G82.65-2.00 &   82.74 &  -1.98 &  20 53 10.6 &   +41 34 59 &   73  & 0.89 (3) &     F &  LDN914 \\
      G86.97-4.06 &   87.08 &  -4.08 &  21 17 20.6 &   +43 23 48 &   37  & 1.87 (3) &     X &  LDN943, LDN944 \\
      G89.65-7.02 &   89.77 &  -6.83 &  21 38 26.8 &   +43 16 24 &   61  & 1.21 (3) &   FBX &  B159, LDN977 \\
      G93.21+9.55 &   93.20 &   9.56 &  20 37 00.0 &   +56 54 50 &   37  & 0.30 (2) &     X &  LDN1033 \\
      G94.15+6.50 &   94.11 &   6.30 &  20 59 26.2 &   +55 35 58 &   50  & 0.25 (2) &   FBX &  B357 \\
      G98.00+8.75 &   97.87 &   8.72 &  21 03 57.1 &   +59 59 52 &   43  & 1.12 (3) &    FX &  ArchG097.82+08.67 \\
    G105.57+10.39 &  105.60 &  10.30 &  21 41 40.2 &   +66 33 29 &   55  & 0.88 (3) &    FX &  ArchG105.55+10.45 \\
     G107.20+5.52 &  107.20 &   5.52 &  22 21 17.6 &   +63 42 25 &   68  & 0.80 (6) &     X &  PCC249, S9, LDN1204, S140 \\
    G108.28+16.68 &  108.28 &  16.68 &  21 09 52.3 &   +72 53 00 &   37  & -        &     B &   \\
    G109.18-37.59 &  109.18 & -37.59 &  00 03 50.7 &   +24 00 23 &   37  & 0.16 (2) &     I &   \\
      G109.80+2.70  &  109.80 &   2.70 &  22 53 31.3 &   +62 31 44 & 37  & 0.80 (6) &     X &  PCC288, S8 \\
     G110.89-2.78 &  110.86 &  -2.61 &  23 18 20.3 &   +58 04 00 &   49  & -        &    FX &   \\
     G111.41-2.95 &  111.27 &  -3.01 &  23 22 20.1 &   +57 49 55 &   49  & -        &     X &  ArchG111.11-03.01 \\
    G126.63+24.55 &  126.63 &  24.55 &  04 23 49.6 &   +85 47 17 &   37  & 0.15 (6) &   ICF &  S1, LDN1320 \\
     G127.79+2.66 &  127.79 &   2.66 &  01 37 48.0 &   +65 05 30 &   37  & 1.06 (3) &     X &  ArchG127.69+02.65 \\
    G130.37+11.26 &  130.37 &  11.26 &  02 32 15.8 &   +72 39 18 &   37  & 0.81 (3) &    XC &  LDN1340 \\
    G130.42-47.07 &  130.42 & -47.07 &  01 12 34.0 &   +15 29 47 &   37  & 0.34 (2) &     I &   \\
     G131.65+9.75 &  131.65 &   9.75 &  02 39 24.0 &   +70 46 27 &   37  & 0.20 (6) &    CX &  S3,  \\
     G132.12+8.95 &  132.09 &   8.89 &  02 39 54.8 &   +69 48 35 &   43  & 1.10 (3) &     X &   \\
     G149.67+3.56 &  149.54 &   3.45 &  04 18 08.3 &   +55 15 49 &   49  & 0.66 (3) &    FX &  LDN1400, LDN1394, \\
     & & & & & & & & B8, B9, MBL71 \\
     G150.47+3.93 &  150.35 &   3.95 &  04 24 36.8 &   +55 02 00 &   43  & 0.17 (6) &     X &  LDN1399, MLB72, MLB74, \\
     & & & & & & & & LM17, ArchG150.41+03.91 \\
     G151.45+3.95 &  151.45 &   3.95 &  04 29 53.6 &   +54 14 51 &   37  & 0.17 (6) &     I &  B12, LDN1407, LDN1400F, \\
     & & & & & & & & MLB77, LM25 \\
     G154.08+5.23 &  154.06 &   5.15 &  04 47 35.0 &   +53 05 54 &   37  & 0.17 (6) &     I &  LDN1426, LM56 \\
     G157.08-8.68 &  157.08 &  -8.68 &  04 01 41.0 &   +41 14 40 &   44  & 0.35 (5) &    FX &  (LDN1443) \\
     G157.92-2.28 &  157.85 &  -2.35 &  04 28 52.3 &   +45 15 56 &   49  & 1.88 (3) &    XF &   \\
    G159.34+11.21 &  159.51 &  11.30 &  05 41 18.0 &   +52 12 13 &   49  & 0.75 (3) &     F &   \\
     G161.55-9.30 &  161.55 &  -9.30 &  04 16 10.9 &   +37 46 26 &   37  & 0.35 (5) &    FB &  S7 \\
     G163.82-8.44 &  163.82 &  -8.44 &  04 27 10.1 &   +36 46 02 &  100  & 0.35 (5) &     F &   \\
     G164.71-5.64 &  164.60 &  -5.51 &  04 40 43.1 &   +38 09 50 &   75  & 0.14 (6) &     F &  LDN1481 \\
     G167.20-8.69 &  166.99 &  -8.71 &  04 36 34.3 &   +34 16 34 &   55  & 0.35 (5) &    FX &   \\
    G168.85-10.19 &  168.92 & -10.24 &  04 37 04.8 &   +31 49 48 &   37  & 0.35 (5) &     I &   \\
     G173.43-5.44 &  173.52 &  -5.27 &  05 08 42.3 &   +31 23 17 &   61  & 1.06 (3) &     X &   \\
     G176.27-2.09 &  176.27 &  -2.09 &  05 28 14.0 &   +30 57 36 &   37  & 2.00 (6) &     F &  S6 \\
    G181.84-18.46 &  181.84 & -18.46 &  04 43 56.0 &   +16 57 27 &   37  & 0.35 (5) &     F &   \\
    G188.24-12.97 &  188.33 & -13.11 &  05 17 05.2 &   +14 54 58 &   61  & -        &    FX &   \\
    G189.51-10.41 &  189.58 & -10.21 &  05 29 54.7 &   +15 26 45 &   55  & -        &    FX &   \\
     G195.74-2.29 &  195.74 &  -2.29 &  06 10 58.3 &   +14 09 30 &   37  & 1.09 (3) &     I &  ArchG195.73-02.39 \\
     G198.58-9.10 &  198.65 &  -9.10 &  05 52 29.0 &   +08 19 05 &   49  & 0.90 (6) &    FB &  LDN1598 \\
     G202.23-3.38 &  202.13 &  -3.35 &  06 19 33.4 &   +08 02 28 &   37  & -        &    FI &   \\
     G203.42-8.29 &  203.62 &  -8.36 &  06 04 35.9 &   +04 22 03 &   49  & 0.34 (2) &   FBX &   \\
     G205.06-6.04 &  205.21 &  -5.85 &  06 16 26.2 &   +04 09 26 &   55  & 0.45 (5) &    FX &   \\
    G210.90-36.55 &  210.90 & -36.56 &  04 35 07.0 &   -14 15 08 &   73  & 0.11 (6) &    FI &  LDN1642, MBM20, IREC305 \\
    G212.07-15.21 &  212.16 & -15.13 &  05 55 57.7 &   -06 09 57 &   37  & 0.45 (5) &    FX &   \\
     G215.37-3.04 &  215.37 &  -3.04 &  06 45 03.3 &   -03 33 10 &   37  & 2.77 (3) &     X &   \\
    G215.44-16.38 &  215.44 & -16.38 &  05 57 02.3 &   -09 33 22 &   37  & 0.45 (5) &     F &  S4 \\
     G216.76-2.58 &  216.76 &  -2.64 &  06 49 00.0 &   -04 36 43 &   37  & 2.32 (3) &     X &   \\
     G218.06+2.12 &  218.06 &   2.12 &  07 08 21.6 &   -03 35 33 &   37  & -        &     I &   \\
     G227.95-2.98 &  227.93 &  -3.15 &  07 07 47.0 &   -14 46 45 &   55  & -        &     F &   \\
     G276.78+1.75 &  276.78 &   1.75 &  09 50 21.0 &   -51 40 51 &   77  & 2.00 (6) &     F &  S5, FeSt2-72, DCld 276.9+01.7, \\
     & & & & & & & & DCld 276.8+0.1.9 \\
    G298.31-13.05 &  298.28 & -13.08 &  11 39 00.0 &   -75 17 27 &   37  & 0.15 (5) &     F &  SDN138, FeSt2-129, \\
    & & & & & & & & HMSTG298.3-13.1, FeSt1-188, \\
    & & & & & & & & DCld 298.3-13.1 \\
     G300.61-3.13 &  300.63 &  -3.02 &  12 28 57.2 &   -65 47 16 &   43  & -        &    FX &  HMSTG300.6-3.0 \\
     G300.86-9.00 &  300.86 &  -9.00 &  12 25 16.5 &   -71 46 03 &   37  & 0.23 (6) &     F &  PCC550, S10, SDN143, VMF32, \\
           & & & & & & & & Musca DN Complex \\
    G315.88-21.44 &  315.88 & -21.44 &  17 19 39.2 &   -76 54 40 &   37  & -        &    FB &   \\
     G334.65+2.67 &  334.76 &   2.72 &  16 14 45.9 &   -47 11 32 &   43  & -        &     X &   \\
     G339.22-6.02 &  339.22 &  -6.02 &  17 12 09.3 &   -49 34 29 &   37  & -        &    FX &   \\
     G343.64-2.31 &  343.76 &  -2.37 &  17 10 29.0 &   -43 46 00 &   49  & 1.10 (3) &     X &  HMSTG343.7-2.3, (FeSt1-373) \\
    G358.96+36.75 &  358.96 &  36.75 &  15 39 38.0 &   -07 13 10 &   37  & 0.11 (6) &     I &  LDN1780, LDN1778, MBM33 \\
\end{longtable}

\clearpage

\begin{landscape}
\begin{longtable}{llllllllll}
\caption{Masses of the mapped fields and of the selected clumps. The
columns are (1) the name of the field, (2) the total mass of the mapped area,
(3)-(5) coordinates and  radius of a selected aperture, (6)-(8) temperature,
peak column density, and mass within the aperture as
derived from the total surface brightness without the subtraction of the
local background, 
(9)-(10) temperature and mass based aperture photometry with a local 
background subtraction.
} 
\label{table:mass} \\ \hline \hline
Name  &   $M_{\rm cloud}$
      &   $\alpha$        &  $\delta$          &  Radius         
      &   $T^{\rm map}$   &  $M^{\rm map}$   &  $N^{\rm map}_{\rm max}$  &  $T^{\rm bgsub}$  &  $M^{\rm bgsub}$ \\
      &   ($M_{\sun}$)
      &   (J2000)          &  (J2000)            & ($\arcmin$)        
      &   (K)              &   ($M_{\sun}$)    &  ($10^{21}$\,cm$^{-2}$)    &  (K)               &  ($M_{\sun}$) \\
\hline
\endfirsthead
\caption{continued.} \\
\hline \hline
Name  &   $M_{\rm cloud}$
      &   $\alpha$        &  $\delta$          &  Radius         
      &   $T^{\rm map}$   &  $M^{\rm map}$   &  $N^{\rm map}_{\rm max}$  &  $T^{\rm bgsub}$  &  $M^{\rm bgsub}$ \\
      &   ($M_{\sun}$)
      &   (J2000)          &  (J2000)            & ($\arcmin$)        
      &   (K)              &  ($M_{\sun}$)      &  ($10^{21}$\,cm$^{-2}$)   &  (K)               &  ($M_{\sun}$) \\
\hline
\endhead
\hline
\endfoot
G1.94+6.07         &     137(    11)       &   17 28 14.5  &    -24 11 26  &   4  & 16.3( 0.2)  &  5.6( 0.3)    &    3.8( 0.5)   &  15.7( 0.3)   &  1.2( 0.1)  \\
                   &                       &   17 28 46.2  &    -23 53 16  &   3  & 15.9( 0.2)  &  3.4( 0.2)    &    4.7( 0.8)   &  14.0( 0.3)   &  1.1( 0.1)  \\
G4.18+35.79        &    68.5(  18.8)       &   15 53 30.1  &    -04 39 49  &   4  & 12.8( 0.5)  & 14.0( 2.1)    &     19(   5)   &  11.2( 0.1)   &  8.0( 0.5)  \\
                   &                       &   15 53 35.7  &    -04 47 10  &   3  & 13.6( 0.4)  &  4.9( 0.5)    &    9.0( 2.2)   &  11.8( 0.5)   &  1.9( 0.3)  \\
G6.03+36.73        &    49.5(  13.0)       &   15 54 10.8  &    -02 50 24  &   4  & 11.6( 1.0)  & 14.9( 4.1)    &     58(  22)   &   9.6( 0.1)   & 13.2( 0.7)  \\
                   &                       &   15 54 11.5  &    -02 58 58  &   2  & 12.4( 0.6)  &  2.1( 0.5)    &     14(   4)   &  10.7( 0.4)   &  1.2( 0.2)  \\
G21.26+12.11       &    1270(   130)       &   17 46 53.7  &    -04 34 24  &   3  & 14.0( 0.7)  &  113(  25)    &     18(   4)   &  12.1( 0.3)   &   69(   6)  \\
                   &                       &   17 46 26.1  &    -04 42 55  &   2  & 14.9( 0.2)  &   29(   3)    &    4.1( 0.7)   &  14.2( 0.4)   & 10.4( 0.8)  \\
G26.34+8.65        &    2680(   160)       &   18 08 35.7  &    -01 49 47  &   3  & 15.0( 0.7)  &  131(  25)    &     10(   2)   &  12.7( 0.2)   &   63(   4)  \\
                   &                       &   18 08 39.0  &    -01 57 16  &   2  & 15.5( 0.2)  &   50(   4)    &    4.0( 0.7)   &  14.4( 0.5)   &  9.8( 1.1)  \\
G39.65+1.75        &   50260(  3970)       &   18 57 30.0  &    +06 47 37  &   4  & 15.3( 0.6)  & 3119( 391)    &     27(   6)   &  12.9( 0.2)   & 1251(  60)  \\
                   &                       &   18 56 58.2  &    +07 00 29  &   5  & 16.5( 1.2)  & 4152( 591)    &     28(   3)   &  17.0( 0.2)   & 1237(  50)  \\
G62.16-2.92        &    3800(   160)       &   19 59 44.4  &    +24 16 39  &   5  & 16.9( 0.1)  &  377(  18)    &    3.0( 0.4)   &  15.5( 0.3)   &   58(   4)  \\
G69.57-1.74        &   17490(  1540)       &   20 13 05.9  &    +31 18 16  &   2  & 18.0( 1.0)  &  310(  39)    &     20(   4)   &  13.8( 0.5)   &  146(  19)  \\
                   &                       &   20 13 24.4  &    +31 05 37  &   2  & 17.1( 0.8)  &  244(  26)    &    9.6( 2.0)   &  12.9( 0.3)   &  153(  13)  \\
G82.65-2.00        &   28100(  2510)       &   20 51 34.6  &    +41 26 33  &   4  & 13.9( 1.2)  &  764( 188)    &     48(  21)   &   9.0( 0.3)   &  539(  84)  \\
                   &                       &   20 53 57.7  &    +41 44 20  &   3  & 14.0( 0.7)  &  406(  53)    &     32(   9)   &  11.0( 0.3)   &  222(  25)  \\
G86.97-4.06        &    7770(   790)       &   21 17 42.0  &    +43 19 20  &   3  & 15.4( 0.8)  &  480(  75)    &     14(   3)   &  13.2( 0.3)   &  260(  21)  \\
                   &                       &   21 17 48.3  &    +43 28 13  &   2  & 15.8( 0.4)  &  176(  15)    &    4.0( 0.8)   &  13.5( 0.5)   &   48(   8)  \\
G89.65-7.02        &    8670(  1540)       &   21 37 10.6  &    +43 21 41  &   3  & 14.2( 0.6)  &  267(  72)    &     21(   4)   &  12.9( 0.2)   &  173(  10)  \\
                   &                       &   21 38 33.3  &    +43 13 30  &   2  & 13.8( 0.4)  &  110(  17)    &    7.1( 1.6)   &  12.6( 0.2)   &   47(   2)  \\
G93.21+9.55        &     249(    34)       &   20 37 19.6  &    +56 54 49  &   3  & 14.0( 0.8)  & 17.7( 2.6)    &     24(   6)   &  12.0( 0.3)   & 10.7( 0.9)  \\
                   &                       &   20 36 21.6  &    +56 51 55  &   2  & 14.1( 0.6)  &  6.1( 1.3)    &    7.5( 1.7)   &  12.3( 0.6)   &  3.3( 0.7)  \\
G94.15+6.50        &     346(    38)       &   20 58 39.6  &    +55 44 46  &   3  & 14.6( 0.6)  &  9.0( 1.6)    &    6.9( 1.5)   &  12.7( 0.2)   &  4.2( 0.3)  \\
                   &                       &   20 59 13.1  &    +55 35 10  &   3  & 14.4( 0.6)  & 10.4( 1.4)    &    9.1( 2.5)   &  11.4( 0.7)   &  4.0( 0.8)  \\
G98.00+8.75        &    5690(   410)       &   21 04 21.7  &    +60 09 06  &   2  & 14.6( 0.6)  &   95(   9)    &     13(   3)   &  11.7( 0.5)   &   43(   7)  \\
                   &                       &   21 03 11.5  &    +60 00 55  &   2  & 15.1( 0.4)  &   72(   6)    &    6.6( 1.4)   &  13.3( 0.6)   & 19.6( 3.3)  \\
G105.57+10.39      &    4730(   630)       &   21 43 28.3  &    +66 37 03  &   2  & 15.6( 0.4)  &   29(   3)    &    4.5( 1.6)   &  12.4( 3.8)   &  4.0( 3.9)  \\
                   &                       &   21 40 45.5  &    +66 25 41  &   2  & 14.8( 0.6)  &   58(   9)    &     12(   3)   &  12.2( 0.4)   &   22(   3)  \\
G107.20+5.52 (PCC249)  &   15370(  2450)       &   22 21 25.8  &    +63 51 57  &   3  & 15.9( 0.9)  &  327(  47)    &     74(  11)   &  15.8( 0.2)   &  175(   7)  \\
                   &                       &   22 19 35.1  &    +63 33 45  &   3  & 16.0( 0.8)  &  371(  58)    &     58(   9)   &  14.7( 0.4)   &  190(  14)  \\
G109.18-37.59      &     9.3(   1.4)       &   00 03 52.1  &    +24 00 30  &   5  & 16.4( 0.5)  &  1.3( 0.1)    &   0.78(0.11)   &  16.1( 0.2)   &  0.4( 0.0)  \\
G109.80+2.70 (PCC288)    &    4230(   450)       &   22 53 37.1  &    +62 32 06  &   2  & 14.1( 0.5)  &  163(  29)    &     47(  10)   &  12.6( 0.3)   &   95(   8)  \\
                   &                       &   22 53 58.2  &    +62 22 36  &   3  & 17.6( 0.8)  &  217(  29)    &     14(   2)   &  18.3( 1.3)   &   55(   9)  \\
G126.63+24.55      &    12.0(   1.7)       &   04 23 52.2  &    +85 47 55  &   4  & 14.8( 0.6)  &  1.8( 0.5)    &    3.1( 0.6)   &  14.2( 0.2)   &  1.4( 0.1)  \\
                   &                       &   04 18 27.6  &    +85 42 53  &   2  & 15.3( 0.5)  &  0.3( 0.1)    &    1.4( 0.3)   &  14.7( 0.5)   &  0.2( 0.0)  \\
G127.79+2.66       &    4300(   350)       &   01 37 33.1  &    +65 07 45  &   2  & 14.3( 0.3)  &   83(   7)    &    6.8( 1.6)   &  12.5( 0.3)   &   25(   2)  \\
                   &                       &   01 38 47.3  &    +65 04 24  &   4  & 13.9( 0.6)  &  443(  64)    &     17(   4)   &  12.6( 0.1)   &  244(   9)  \\
G130.37+11.26      &    2730(   430)       &   02 31 59.6  &    +72 39 35  &   4  & 12.6( 0.7)  &  404(  80)    &     25(   7)   &  11.1( 0.1)   &  279(  11)  \\
                   &                       &   02 33 32.3  &    +72 47 44  &   2  & 13.4( 0.4)  &   48(   5)    &    8.7( 2.7)   &  11.3( 1.1)   & 11.8( 4.6)  \\
G130.42-47.07      &    19.0(   4.2)       &   01 12 54.3  &    +15 28 17  &   5  & 16.8( 1.3)  &  2.4( 0.3)    &   0.38(0.07)   &  14.6( 0.7)   &  0.6( 0.1)  \\
G131.65+9.75       &    95.5(  13.3)       &   02 40 12.3  &    +70 36 09  &   2  & 13.9( 0.7)  &  2.3( 0.4)    &     12(   3)   &  12.4( 0.9)   &  1.3( 0.3)  \\
                   &                       &   02 39 56.6  &    +70 42 13  &   3  & 13.0( 0.6)  &  9.7( 2.1)    &     18(   5)   &  11.5( 0.2)   &  6.7( 0.5)  \\
G132.12+8.95       &    5490(   700)       &   02 41 50.5  &    +69 52 44  &   3  & 13.6( 0.8)  &  324(  64)    &     18(   4)   &  12.2( 0.2)   &  150(   7)  \\
                   &                       &   02 40 26.4  &    +69 49 53  &   2  & 13.0( 0.8)  &  145(  38)    &     17(   6)   &  10.4( 0.4)   &  100(  20)  \\
G149.67+3.56       &    3290(   530)       &   04 18 22.5  &    +55 14 15  &   4  & 14.1( 0.4)  &  186(  28)    &    8.8( 2.0)   &  12.4( 0.4)   &   59(   8)  \\
                   &                       &   04 16 45.7  &    +55 10 34  &   2  & 14.7( 0.1)  &   31(   2)    &    4.6( 0.9)   &  14.2( 1.2)   &  3.0( 0.7)  \\
G150.47+3.93       &     230(    44)       &   04 24 31.7  &    +55 02 46  &   3  & 12.9( 0.7)  & 11.9( 1.9)    &     23(   7)   &  11.0( 0.4)   &  5.9( 0.7)  \\
                   &                       &   04 23 55.0  &    +54 57 53  &   2  & 13.0( 0.5)  &  4.6( 0.8)    &     15(   4)   &  11.0( 0.4)   &  1.8( 0.3)  \\
G151.45+3.95       &     120(    18)       &   04 29 53.7  &    +54 15 39  &   4  & 14.0( 0.2)  & 16.0( 1.9)    &    9.8( 2.0)   &  13.0( 0.1)   &  7.3( 0.3)  \\
                   &                       &   04 29 16.9  &    +54 07 28  &   2  & 14.0( 0.5)  &  3.1( 0.5)    &     11(   2)   &  12.5( 0.6)   &  1.6( 0.3)  \\
G154.08+5.23       &    87.9(   7.2)       &   04 48 06.6  &    +53 08 10  &   4  & 13.6( 0.8)  & 10.7( 2.5)    &     20(   5)   &  11.6( 0.1)   &  7.3( 0.3)  \\
                   &                       &   04 47 13.8  &    +53 01 13  &   2  & 13.6( 0.5)  &  2.5( 0.3)    &    8.3( 2.0)   &  12.2( 0.7)   &  1.0( 0.2)  \\
G157.08-8.68       &     688(   132)       &   04 01 41.5  &    +41 04 11  &   4  & 13.4( 0.5)  &   56(   8)    &     29(   7)   &  11.7( 0.3)   &   21(   2)  \\
                   &                       &   04 02 08.9  &    +41 22 21  &   2  & 13.7( 0.5)  & 10.4( 1.3)    &     11(   3)   &  11.6( 0.5)   &  4.1( 0.9)  \\
G157.92-2.28       &   20800(  2150)       &   04 28 40.8  &    +45 05 20  &   3  & 14.1( 0.4)  &  545(  59)    &    8.6( 2.0)   &  12.2( 0.2)   &  187(   9)  \\
                   &                       &   04 28 43.2  &    +45 24 01  &   2  & 13.9( 0.6)  &  193(  25)    &    7.4( 2.7)   &  10.6( 0.9)   &   65(  23)  \\
G159.34+11.21      &    1600(   320)       &   05 42 14.3  &    +52 08 14  &   4  & 14.4( 0.2)  &  125(  18)    &    5.1( 1.0)   &  13.6( 0.2)   &   57(   3)  \\
                   &                       &   05 40 12.1  &    +52 17 15  &   2  & 14.6( 0.2)  &   26(   3)    &    3.5( 0.7)   &  13.8( 0.5)   &  8.9( 1.3)  \\
G161.55-9.30       &     392(    81)       &   04 16 54.7  &    +37 45 26  &   3  & 14.8( 0.2)  & 19.4( 1.6)    &    5.2( 1.0)   &  13.7( 0.2)   &  6.4( 0.4)  \\
                   &                       &   04 16 17.4  &    +37 45 06  &   3  & 14.6( 0.5)  &   20(   3)    &    8.0( 1.7)   &  12.6( 0.4)   &  7.9( 1.0)  \\
G163.82-8.44       &    3500(   560)       &   04 27 03.8  &    +36 58 40  &   5  & 13.9( 0.6)  &   72(  12)    &     17(   4)   &  12.3( 0.2)   &   33(   2)  \\
                   &                       &   04 29 10.8  &    +36 46 35  &   4  & 14.0( 0.3)  &   43(   4)    &    9.2( 1.9)   &  12.8( 0.3)   & 12.9( 1.3)  \\
G164.71-5.64       &     237(    28)       &   04 41 12.6  &    +37 45 55  &   5  & 14.6( 0.2)  &  8.0( 0.7)    &    5.4( 1.0)   &  13.4( 0.2)   &  2.1( 0.1)  \\
                   &                       &   04 40 36.3  &    +37 59 33  &   3  & 14.5( 0.2)  &  3.0( 0.2)    &    4.9( 1.1)   &  12.6( 0.6)   &  0.6( 0.1)  \\
G167.20-8.69       &     566(    55)       &   04 37 21.6  &    +34 04 57  &   4  & 15.0( 0.2)  & 18.2( 1.4)    &    3.2( 0.6)   &  14.1( 0.5)   &  4.4( 0.6)  \\
                   &                       &   04 36 07.9  &    +34 24 17  &   4  & 14.8( 0.3)  & 19.1( 2.2)    &    4.4( 0.9)   &  13.4( 0.5)   &  4.7( 0.6)  \\
G168.85-10.19      &     192(    17)       &   04 37 08.1  &    +31 43 10  &   3  & 14.5( 0.2)  &  8.2( 0.7)    &    2.3( 0.4)   &  13.6( 0.2)   &  2.5( 0.1)  \\
                   &                       &   04 37 10.4  &    +31 56 30  &   4  & 14.6( 0.2)  & 14.4( 1.7)    &    2.4( 0.4)   &  13.9( 0.4)   &  3.5( 0.4)  \\
G173.43-5.44       &    5640(   390)       &   05 09 34.5  &    +31 08 24  &   3  & 14.7( 0.3)  &   83(  13)    &    2.5( 0.5)   &  13.2( 0.3)   &   30(   3)  \\
                   &                       &   05 08 01.8  &    +31 39 55  &   3  & 15.1( 0.2)  &   88(   6)    &    2.4( 0.4)   &  14.5( 0.6)   & 13.8( 2.0)  \\
G176.27-2.09       &   14720(  1120)       &   05 28 50.9  &    +30 51 28  &   2  & 13.5( 0.6)  &  429(  75)    &     17(   4)   &  11.8( 0.5)   &  227(  33)  \\
                   &                       &   05 28 00.0  &    +31 01 13  &   2  & 13.2( 0.7)  &  438(  75)    &     14(   4)   &  11.3( 0.3)   &  258(  25)  \\
G181.84-18.46      &     230(    32)       &   04 43 41.0  &    +16 56 27  &   3  & 14.6( 0.3)  & 14.6( 1.6)    &    5.6( 1.1)   &  13.4( 0.4)   &  5.0( 0.5)  \\
                   &                       &   04 44 44.0  &    +17 00 22  &   3  & 14.9( 0.2)  & 11.3( 1.8)    &    3.4( 0.7)   &  13.3( 0.9)   &  2.6( 0.6)  \\
G195.74-2.29       &    5030(   370)       &   06 10 51.7  &    +14 10 10  &   4  & 14.7( 0.9)  &  588(  76)    &     49(  11)   &  12.5( 0.1)   &  337(   9)  \\
                   &                       &   06 11 29.6  &    +14 05 39  &   2  & 14.3( 0.3)  &  109(   8)    &    9.7( 2.0)   &  13.0( 0.5)   &   37(   5)  \\
G198.58-9.10       &    3230(   420)       &   05 52 26.5  &    +08 13 25  &   4  & 14.3( 1.1)  &  313(  79)    &     28(   6)   &  12.4( 0.2)   &  237(  13)  \\
                   &                       &   05 52 59.4  &    +08 23 07  &   2  & 14.8( 0.3)  &   43(   4)    &    3.9( 0.7)   &  13.8( 0.7)   & 12.1( 2.0)  \\
G203.42-8.29       &     402(    68)       &   06 04 26.2  &    +04 11 37  &   3  & 13.9( 0.7)  & 16.6( 3.6)    &    9.3( 2.2)   &  12.1( 0.3)   & 10.8( 1.1)  \\
                   &                       &   06 05 05.2  &    +04 23 09  &   2  & 14.7( 0.4)  &  5.8( 0.6)    &    4.9( 1.2)   &  12.3( 0.4)   &  1.6( 0.2)  \\
G205.06-6.04       &     886(   100)       &   06 15 28.9  &    +04 12 23  &   4  & 15.0( 0.3)  &   38(   4)    &    4.9( 0.9)   &  14.0( 0.3)   & 15.4( 1.0)  \\
                   &                       &   06 16 04.5  &    +04 01 00  &   4  & 14.7( 0.5)  &   40(   7)    &    6.8( 1.3)   &  13.5( 0.3)   & 16.6( 1.5)  \\
G210.90-36.55      &    64.7(   8.1)       &   04 35 08.9  &    -14 14 48  &   7  & 13.9( 0.5)  &  9.3( 2.0)    &    8.8( 1.9)   &  12.7( 0.1)   &  6.1( 0.2)  \\
                   &                       &   04 34 00.5  &    -14 10 39  &   7  & 15.2( 0.3)  &  4.0( 0.5)    &    2.2( 0.4)   &  14.5( 0.2)   &  1.5( 0.0)  \\
G212.07-15.21      &     315(    38)       &   05 55 22.7  &    -06 11 18  &   3  & 15.8( 0.3)  & 13.2( 1.0)    &    2.0( 0.4)   &  13.8( 0.7)   &  2.2( 0.4)  \\
                   &                       &   05 56 21.0  &    -06 12 19  &   2  & 15.5( 0.2)  &  5.4( 0.4)    &    1.7( 0.4)   &  14.0( 0.8)   &  1.2( 0.2)  \\
G215.37-3.04       &   28180(  2300)       &   06 45 05.7  &    -03 37 40  &   2  & 13.7( 0.3)  &  477(  29)    &    5.2(12.4)   &   8.4( 3.9)   &   77(  82)  \\
                   &                       &   06 45 30.4  &    -03 33 28  &   2  & 13.6( 0.2)  &  493(  28)    &    4.9( 1.8)   &  10.9( 0.7)   &   97(  33)  \\
G215.44-16.38      &     379(    40)       &   05 57 13.6  &    -09 35 50  &   2  & 14.2( 0.4)  & 12.6( 2.0)    &    7.9( 1.6)   &  13.1( 0.4)   &  6.3( 0.7)  \\
                   &                       &   05 56 32.6  &    -09 30 31  &   2  & 14.9( 0.2)  &  8.2( 0.6)    &    3.2( 0.6)   &  14.4( 1.1)   &  1.1( 0.3)  \\
G216.76-2.58       &   29940(  2730)       &   06 49 00.3  &    -04 37 01  &   2  & 12.2( 0.4)  &  833( 115)    &     17(   6)   &  10.0( 0.5)   &  355(  64)  \\
                   &                       &   06 48 39.9  &    -04 38 36  &   2  & 12.3( 0.4)  &  753(  80)    &     15(   6)   &   9.9( 0.4)   &  254(  51)  \\
G276.78+1.75       &   40100(  2560)       &   09 49 28.5  &    -51 23 00  &   5  & 14.4( 0.2)  & 1006(  83)    &    4.4( 0.9)   &  13.0( 0.2)   &  226(  14)  \\
                   &                       &   09 52 03.2  &    -51 56 51  &   3  & 14.6( 0.3)  &  322(  25)    &    4.6( 1.1)   &  12.4( 0.4)   &   67(  10)  \\
G298.31-13.05      &    34.1(   4.0)       &   11 37 58.0  &    -75 21 26  &   3  & 14.1( 0.3)  &  2.6( 0.5)    &    6.1( 1.2)   &  13.3( 0.4)   &  1.5( 0.1)  \\
                   &                       &   11 40 09.0  &    -75 15 23  &   3  & 13.9( 0.6)  &  2.8( 0.5)    &    5.7( 1.3)   &  12.6( 0.2)   &  1.7( 0.1)  \\
G300.86-9.00 (PCC550)   &     128(    23)       &   12 24 32.8  &    -71 50 00  &   3  & 13.9( 0.4)  & 10.5( 2.5)    &    9.1( 1.9)   &  12.8( 0.4)   &  5.6( 0.6)  \\
                   &                       &   12 25 34.9  &    -71 42 14  &   3  & 13.8( 0.3)  & 11.4( 1.6)    &    9.2( 2.0)   &  12.7( 0.2)   &  6.0( 0.4)  \\
G343.64-2.31       &   15840(  1910)       &   17 11 03.2  &    -43 40 11  &   3  & 15.4( 0.9)  &  621( 106)    &     27(   5)   &  13.5( 0.2)   &  281(  12)  \\
                   &                       &   17 09 21.0  &    -43 46 38  &   3  & 17.2( 0.5)  &  393(  39)    &     10(   2)   &  14.9( 0.6)   &   83(  11)  \\
G358.96+36.75      &    15.3(   4.2)       &   15 39 33.1  &    -07 10 47  &   5  & 15.3( 0.3)  &  3.2( 0.5)    &    4.1( 0.7)   &  14.5( 0.2)   &  1.6( 0.1)  \\
\end{longtable}
\end{landscape}

\clearpage

\begin{longtable}{lccccccccccc}
\caption{Properties of selected filaments. The columns are: 
(1) name of the field, 
(2) the length of the analysed filament, 
(3) width of the analysed filament area, 
(4) total mass of this area, 
(5)--(6) average column density along the filament and its dispersion, 
(7) Jeans length calculated for average and peak column density, 
(8) measured FWHM of the mean profile of the filament, 
(9)-(11) parameters of the Plummer profile fit,
(12) mass per unit length according to the fit of the Plummer function.
} 
\label{table:filaments} \\
\hline\hline            
Name &   Length           &  Width                &    Mass   
     &   $N_{\rm H_2}$           &  $\sigma(N(H_{2})$
     &   $L_{\rm Jeans}$  
     &   FWHM             &  $\rho_{\rm c}$     &  $R_{\rm flat}$    &  $p$   
     &   $M_{\rm line}$   \\
     &   (pc)             &  (pc)                 &  ($M_{\sun}$)
     &   \multicolumn{2}{c}{($10^{21}$/cm$^{2}$)}
     &   (pc)             
     &   (pc)             &   (H$_{2}$/cm$^3$)    &  (pc)               &
     &   ($M_{\sun}$/pc)  \\
\hline                       
\endfirsthead
\caption{continued.} \\
\hline \hline
Name &   Length           &  Width                &    Mass   
     &   $N_{\rm H_2}$           &  $\sigma(N(H_{2})$
     &   $L_{\rm Jeans}$  
     &   FWHM             &  $\rho_{\rm c}$     &  $R_{\rm flat}$    &  $p$   
     &   $M_{\rm line}$   \\
     &   (pc)             &  (pc)                 &  ($M_{\sun}$)
     &   \multicolumn{2}{c}{($10^{21}$/cm$^{2}$)}
     &   (pc)    
     &   (pc)             &   (H$_{2}$/cm$^3$)    &  (pc)               &
     &   ($M_{\sun}$/pc)  \\
\hline
\endhead
\hline
\endfoot
G1.94+6.07          &   2.49  & 0.40 &   19.4  &        1.7 &    0.2  &     0.32/0.24     &     0.21 &  2.41e+03  &   0.06  &   1.5 &      7.0   \\ 
G82.65-2.00         &  14.94  & 0.40 &  2931.7  &       19.2 &    3.9  &     0.03/0.02     &     0.24 &  3.70e+04  &   0.06  &   2.0 &     116.4   \\ 
G89.65-7.02         &  17.83  & 0.80 &  2283.9  &        7.1 &    2.7  &     0.08/0.04     &     0.30 &  6.56e+03  &   0.10  &   1.7 &     61.9   \\ 
G94.15+6.50         &   2.46  & 0.40 &   47.2  &        4.0 &    0.9  &     0.13/0.09     &     0.11 &  1.23e+04  &   0.08  &   6.6 &      7.0   \\ 
G98.00+8.75         &   5.78  & 0.40 &  606.7  &        7.4 &    2.1  &     0.07/0.05     &     0.22 &  5.11e+04  &   0.02  &   1.6 &     42.8   \\ 
G105.57+10.39       &   3.64  & 0.40 &  237.7  &        4.3 &    0.5  &     0.12/0.11     &     0.20 &  1.63e+04  &   0.03  &   1.7 &     24.3   \\ 
G126.63+24.55       &   0.78  & 0.40 &    3.5  &        1.9 &    0.4  &     0.28/0.19     &     0.12 &  3.97e+03  &   0.06  &   2.6 &      7.0   \\ 
G149.67+3.56        &   6.54  & 0.40 &  691.1  &        5.3 &    0.6  &     0.10/0.09     &     0.25 &  4.88e+03  &   0.05  &   1.3 &     42.2   \\ 
G157.08-8.68        &   3.47  & 0.40 &  209.5  &       11.5 &    2.6  &     0.05/0.03     &     0.63 &  3.45e+04  &   0.04  &   2.1 &     51.6   \\ 
G157.92-2.28        &  27.66  & 0.40 &  3151.8  &        3.7 &    1.0  &     0.14/0.09     &     0.35 &  9.85e+03  &   0.04  &   1.5 &     38.9   \\ 
G159.34+11.21       &   6.69  & 0.80 &  307.7  &        2.9 &    0.6  &     0.19/0.13     &     0.31 &  1.37e+03  &   0.15  &   1.4 &     30.0   \\ 
G161.55-9.30        &   2.41  & 0.40 &   92.7  &        4.4 &    0.7  &     0.12/0.09     &     0.65 &  7.36e+03  &   0.05  &   1.6 &     29.1   \\ 
G163.82-8.44        &   8.22  & 0.40 &  344.7  &       11.4 &    9.5  &     0.05/0.01     &     0.65 &  4.03e+04  &   0.03  &   1.9 &     39.4   \\ 
G164.71-5.64        &   2.36  & 0.40 &   23.0  &        2.4 &    0.4  &     0.22/0.16     &     0.16 &  6.10e+03  &   0.02  &   1.4 &      6.3   \\ 
G167.20-8.69        &   3.29  & 0.40 &   51.6  &        2.1 &    0.5  &     0.25/0.18     &     0.67 &  5.14e+03  &   0.03  &   1.5 &      7.9   \\ 
G176.27-2.09        &  13.25  & 0.80 &  1808.6  &        8.6 &    2.7  &     0.06/0.04     &     0.37 &  1.15e+04  &   0.09  &   1.8 &     82.0   \\ 
G181.84-18.46       &   2.70  & 0.40 &   58.1  &        3.3 &    0.7  &     0.16/0.12     &     0.66 &  6.55e+03  &   0.04  &   1.5 &     19.3   \\ 
G198.58-9.10        &   5.78  & 0.40 &  820.2  &       14.9 &    4.7  &     0.03/0.02     &     0.29 &  2.34e+04  &   0.06  &   1.7 &     123.5   \\ 
G203.42-8.29        &   2.87  & 0.40 &   61.6  &        4.0 &    1.1  &     0.13/0.08     &     0.14 &  1.34e+04  &   0.03  &   1.6 &     17.1   \\ 
G205.06-6.04        &   5.46  & 0.40 &  142.2  &        2.2 &    0.5  &     0.24/0.15     &     0.52 &  1.74e+03  &   0.08  &   1.4 &     20.1   \\ 
G210.90-36.55       &   1.34  & 0.40 &   12.9  &        4.8 &    1.4  &     0.11/0.07     &     0.17 &  8.39e+03  &   0.07  &   2.1 &     17.5   \\ 
G212.07-15.21       &   6.23  & 0.40 &   77.3  &        0.9 &    0.1  &     0.59/0.48     &     0.41 &  1.29e+03  &   0.05  &   1.4 &      7.3   \\ 
G215.44-16.38       &   3.31  & 0.40 &   80.7  &        4.6 &    1.7  &     0.12/0.07     &     0.32 &  6.93e+03  &   0.10  &   2.6 &     28.1   \\ 
G276.78+1.75        &  12.91  & 0.80 &  925.6  &        3.0 &    1.1  &     0.18/0.10     &     0.49 &  1.54e+03  &   0.55  &   9.1 &     34.2   \\ 
G298.31-13.05       &   1.11  & 0.40 &   10.8  &        3.8 &    0.5  &     0.14/0.10     &     0.11 &  1.08e+04  &   0.04  &   2.2 &     10.8   \\ 
G300.86-9.00        &   1.74  & 0.40 &   42.6  &        6.3 &    1.0  &     0.09/0.07     &     0.19 &  1.07e+04  &   0.10  &   3.5 &     27.9   \\ 
\end{longtable}


\Online

\begin{appendix}

\section{Maps of observed and derived quantities} \label{sect:appendix_ids}

Figures~\ref{fig:ids_00}--\ref{fig:ids_68} show maps of observed and
derived quantities for the fields other than the one already shown in
Figs.~\ref{fig:ids_1}--\ref{fig:ids_1b}. Included in the figures are the colour
temperature map derived from SPIRE data (frame $a$, 40$\arcsec$
resolution), the 250\,$\mu$m SPIRE surface brightness map (frame $b$,
18$\arcsec$ resolution), the AKARI wide filter maps at 140\,$\mu$m and
90\,$\mu$m (frames $c$ and $e$; convolved to a resolution of one arc
minute), the visual extinction $A_{\rm V}$ derived from 2MASS catalogue
stars (frame $d$; resolution of 2$\arcmin$), and the WISE 22\,$\mu$m
intensity (frame $f$; 12$\arcsec$ resolution). In the cases where
public WISE data were not yet available, frame $f$ shows the
25\,$\mu$m IRAS map (resolution of 4.5$\arcmin$). The respective beam
sizes are indicated in the lower right hand corner of each frame. The
black circles indicate the positions of clumps selected for further
examination in Sect.~\ref{sect:clumps} and the white circles indicates
a reference region used for background subtraction in the SED plots in
Sect.~\ref{sect:clumps}. In frame $a$, the arrow indicates the
position of the stripe plotted in
Appendix~\ref{sect:appendix_stripes}.

\section{Column density maps} \label{sect:appendix_colden}

The figure~\ref{fig:colden_2}--\ref{fig:colden_6} shows the column
density maps for all the fields excluding those already shown in
Fig.~\ref{fig:colden_1}.


\end{appendix}

\begin{appendix}

\section{One-dimensional cuts of the surface brightness data} \label{sect:appendix_stripes}

Figures~\ref{fig:stripes_2}--\ref{fig:stripes_6} show selected surface
brightness values along the one-dimensional cuts marked in the frames
$a$ of Fig.~\ref{fig:ids_1} and
Figs.~\ref{fig:ids_00}--\ref{fig:ids_68}.


\end{appendix}

\begin{appendix}

\section{Figures of the selected elongated cloud structures}
\label{sect:appendix_filaments}

The analysis of the filamentS in the fields G163.82-8.44 and G300.86-9.00
(PCC\,550) were shown in Figs.~\ref{fig:filaments_1}--\ref{fig:filaments_2}. The
plots for the other fields listed in Table~\ref{table:filaments} are shown in
Figs.~\ref{fig:filaments_00}-\ref{fig:filaments_23}. These include only the
fields with an existing distance estimate and where a clear filament or other
distinct elongated structure could be discerned.

\end{appendix}

\begin{appendix}

\section{Spectra of selected regions} \label{sect:appendix_seds}

The spectral energy distributions of selected regions were presented
in Fig.~\ref{fig:sed_1}. Similar figures for the remaining fields are
shown in Figs.~\ref{fig:sed_2}--\ref{fig:sed_6}.


\end{appendix}

\end{document}